\documentclass[sigconf,natbib=true]{acmart}

\AtBeginDocument{%
  \providecommand\BibTeX{{%
    \normalfont B\kern-0.5em{\scshape i\kern-0.25em b}\kern-0.8em\TeX}}}

\usepackage{mkolar_definitions}
\usepackage[ruled, vlined]{algorithm2e}
\usepackage{multirow}

\usepackage{tikz,pgfplots}
\usetikzlibrary{positioning}
\usetikzlibrary{patterns}
\newcommand{\content}[1]{content}
\newcommand{\contents}[1]{contents}
\newcommand{\Content}[1]{Content}
\newcommand{\shortname}{NLB}

\copyrightyear{2024}
\acmYear{2024}
\setcopyright{rightsretained}
\acmConference[WSDM '24]{Proceedings of the 17th ACM International
Conference on Web Search and Data Mining}{March 4--8, 2024}{Merida, Mexico}
\acmBooktitle{Proceedings of the 17th ACM International Conference on Web
Search and Data Mining (WSDM '24), March 4--8, 2024, Merida,
Mexico}\acmDOI{10.1145/3616855.3635833}
\acmISBN{979-8-4007-0371-3/24/03}
\begin{document}

\title{Long-Term Value of Exploration:\\ Measurements, Findings and Algorithms}



\author{Yi Su}
\affiliation{%
  \institution{Google Deepmind}
  \country{}
}
\email{yisumtv@google.com}

\author{Xiangyu Wang}
\affiliation{%
  \institution{Google}
  \country{}
}
\email{xiangyuw@google.com}

\author{Elaine Ya Le}
\affiliation{%
  \institution{Google}
  \country{}
}
\email{elainele@google.com}

\author{Liang Liu}
\affiliation{%
  \institution{Google}
  \country{}
}
\email{liangliu@google.com}

\author{Yuening Li}
\affiliation{%
  \institution{Google}
  \country{}
}
\email{yueningl@google.com}

\author{Haokai Lu}
\affiliation{%
  \institution{Google Deepmind}
  \country{}
}
\email{haokai@google.com}

\author{Benjamin Lipshitz}
\affiliation{%
  \institution{Google}
  \country{}
}
\email{lipshitz@google.com}

\author{Sriraj Badam}
\affiliation{%
  \institution{Google}
  \country{}
}
\email{srirajdutt@google.com}

\author{Lukasz Heldt}
\affiliation{%
  \institution{Google}
  \country{}
}
\email{heldt@google.com}

\author{Shuchao Bi}
\affiliation{%
  \institution{Google}
  \country{}
}
\email{shuchaobi@google.com}

\author{Ed H. Chi}
\affiliation{%
  \institution{Google Deepmind}
  \country{}
}
\email{edchi@google.com}

\author{Cristos Goodrow}
\affiliation{%
  \institution{Google}
  \country{}
}
\email{cristos@google.com}

\author{Su-Lin Wu}
\affiliation{%
  \institution{Google}
  \country{}
}
\email{sulin@google.com}

\author{Lexi Baugher}
\affiliation{%
  \institution{Google}
  \country{}
}
\email{baugher@google.com}

\author{Minmin Chen}
\affiliation{%
  \institution{Google Deepmind}
  \country{}
}
\email{minminc@google.com}

\renewcommand{\shortauthors}{Yi Su, et al.}

\begin{abstract}
Effective exploration is believed to positively influence the long-term user experience on recommendation platforms. Determining its exact benefits, however, has been challenging. Regular A/B tests on exploration often measure neutral or even negative engagement metrics while failing to capture its long-term benefits. We here introduce new experiment designs to formally quantify the long-term value of exploration by examining its effects on content corpus, and connecting content corpus growth to the long-term user experience from real-world experiments.
Once established the values of exploration, we investigate the Neural Linear Bandit algorithm as a general framework to introduce exploration into any deep learning based ranking systems. We conduct live experiments on one of the largest short-form video recommendation platforms that serves billions of users to validate the new experiment designs, quantify the long-term values of exploration, and to verify the effectiveness of the adopted neural linear bandit algorithm for exploration.
\end{abstract}



\maketitle

\section{Introduction}
Recommender systems are becoming ubiquitous in people's daily life, serving users with relevant content on recommendation platforms.
Many of these systems are trained to predict and exploit users' immediate response to recommendations, such as click, dwell time, and purchase, achieving enormous success in personalization~\citep{sarwar2001item, koren2009matrix, covington2016deep, zhang2019deep}.
However, these types of exploitation-based systems are known to suffer from the closed feedback loop effect~\citep{jiang2019degenerate, jadidinejad2020using}, for which the recommender system and the user reinforce each other's choices. Users, presented with the recommended items, provide feedback only on the chosen items; the system, trained with the biased feedback data, further consolidates and reinforces users' profiles towards what they have interacted with before. As a result, users are increasingly confined to a narrower set of content, while a lot of the content on the platform remains undiscovered.

Exploration is the key to breaking such feedback loops. By exposing users to less certain contents
~\citep{jadidinejad2020using, chen2021values},  it actively acquires future learning signals about the unknown user content pairs to fill in the knowledge gap in the system.
Doing so, exploration can introduce users to novel content, which we refer to as user exploration~\cite{chen2021values, song2022show, schnabel2018short}; it can also make more fresh and tail content (as well as content providers) discoverable on the platform, which we refer to as item exploration~\cite{chen2021exploration, aharon2015excuseme}. We focus our discussion on item exploration.
While efficient exploration techniques ~\citep{li2010contextual, agarwal2014taming, silver2016mastering, chen2019top} have been actively studied in the bandits and reinforcement learning literature, deploying them in real-world industrial systems has proven difficult. The main challenge lies in \emph{measuring}~\citep{chen2021exploration} the exact benefit of exploration, which then serves as concrete and measurable evidence to switch from a purely exploitation-based system to an exploration-based one. Although exploration techniques such as Upper-Confidence-Bound~\citep{auer2002finite, chu2011contextual} and Thompson Sampling~\citep{thompson1933likelihood, chapelle2011empirical, riquelme2018deep} have been mathematically proven to attain better regret than greedy ones, it is unclear if this benefit translates to industrial recommendation settings with noisy and delayed feedback, as well as non-testable modeling assumptions. 

There are three main challenges in measuring the benefit of exploration. The first one is the \emph{metric} to be examined, as the benefit of exploration takes a long time to manifest and is hard to be captured in regular A/B testing. Meanwhile, recommending less certain content often leads to short-term user engagement metric loss. Therefore, it is critical to identify some intermediate entities that could serve as proxies connecting exploration to the long-term user experience. 
We focus on studying the value of exploration through its intermediate impact on the content corpus in the system~\footnote{The value of exploration on model learning has been studied previously~\cite{chen2019top, jadidinejad2020using, houthooft2016vime}. We also include experiments and discussions along this line of work in Appendix ~\ref{appsec: model}.}. We present a systematic study of how exploration enlarges the content corpus, which further translates to the long-term user engagement gain.
The second challenge is the \emph{experiment design}, as we will explain later, the commonly used user-diverted A/B testing~\citep{kohavi2020trustworthy} cannot capture the benefit of exploration on the content corpus as the information leak between the control and experiment arms. We introduce a new experiment framework to measure the impact of exploration on the content corpus change. 
The last lies in the \emph{design of the exploration-based systems} that could be served in real-world, industrial-scale setting. 
To this end, we adopt a scalable exploration algorithm, namely neural linear bandits (\shortname) ~\citep{riquelme2018deep}, to fully unlock and examine the potential benefit of exploration. \shortname~ performs linear regression on top of the learned representation from deep neural networks as the contextual features to estimate uncertainty. It fits nicely into the modern deep learning-based recommender systems~\citep{covington2016deep} while attaining simplicity in calculating accurate uncertainty estimates.
In summary, we make the following contributions:
\begin{itemize}
  \item \textbf{Metrics for studying the benefit of exploration}: We bring to light the measurement challenges of exploration and offer the first comprehensive study to systematically quantify the value of exploration in recommender systems. Our approach leverages content corpus as an important intermediate quantity to bridge the gap between exploration and the user experience and define corpus metrics to measure the effectiveness of different exploration treatments.
    \item \textbf{Experiment frameworks for measuring the value of exploration}: To prevent corpus leakage between the control and treatment arm, we propose a new user-corpus-codiverted experiment framework to measure the effect of exploration on the proposed corpus metrics in an unbiased manner. 
    \item \textbf{Designs of exploration-based system through Neural Linear Bandit}: Though the \shortname~ algorithm has been studied theoretically, 
    we discuss the challenges of integrating it into industrial recommender systems and detail our implementations. We further validate its success through large-scale live experiments and point to exciting future directions for building exploration-based recommender systems.
     \item \textbf{Validations and findings through large-scale live experiments}: We validate the experiment designs through extensive live experiments on a large-scale short-form video recommender platform. We present findings on how exploration enlarges content corpus, and ultimately, connect the corpus improvements to long-term user experiences.
\end{itemize}

\section{Related Work}
Our work falls into the general theme of exploration or cold-start in recommender systems. Unlike prior works that mainly focus on algorithmic designs, we study the long-term value of exploration. We then use the Neural Linear Bandit from the online learning literature as the backbone algorithm to build effective exploration-based system. In this section, we discuss the literature along these two lines of research.
\paragraph{Exploration or cold-start in Recommender Systems.}
The problem of exposure bias and its induced closed feedback loop~\citep{jadidinejad2020using} are well-known to the recommender system community. Exploration~\citep{chen2021exploration} is believed to be the key to breaking this closed feedback loop. 
There has been considerable interest in developing new exploration-based systems. \citet{mcinerney2018explore} studied the benefit of $\epsilon$-greedy in balancing exploration with exploitation in production systems, focusing specifically on how exploration techniques could be 
coupled with recommendation explanations. \citet{chen2021values} examined the direct role of exploration on users, i.e., helping them discover new interests and studied its effects on different aspects of the recommendation quality, such as diversity and serendipity. Similarly, \citet{song2022show} studied effective user exploration techniques by introducing a hierarchical bandit framework that facilitates large-space user interest learning. Contrary to these works, we examine the long-term value of exploration through one important but overlooked entity: corpus. We argue that exploration can have a net positive long-term effect on users through this intermediary despite short-term user engagement loss. By designing new experiments and evaluating on a commercial recommendation platform serving billions of users, we showcase the effect of exploration on corpus growth, and further connect the corpus growth with the long-term user experience. Our work provides a strong argument advocating for exploration-based systems. 

\paragraph{Algorithms for Exploration.} Multi-armed bandits and contextual bandits~\citep{lai1985asymptotically, auer2002finite, langford2007epoch, agarwal2014taming} are natural frameworks to study exploration-exploitation trade-offs, which have been widely used in many domains, such as recommender systems~\citep{li2010contextual, joachims2018deep}, healthcare~\citep{durand2018contextual, mintz2020nonstationary}, dynamic pricing~\citep{misra2019dynamic}, and dialogue systems~\citep{liu2018customized}. Among its variants, the most well-studied setting is bandits with linear payoffs~\citep{auer2002finite, chu2011contextual, abbasi2011improved}, with LinUCB~\citep{chu2011contextual} and Thompson Sampling (TS)~\citep{agrawal2013thompson} being the two representative approaches. 
Compared with UCB, TS shows stronger performance empirically~\citep{chapelle2011empirical}. With the success of deep learning models, there has been a surge of interest in exploring contextual bandits with neural network models in recent years. \citet{riquelme2018deep} proposed NeuralLinear algorithm, which performs Bayesian Linear Regression at the last layer of the network. \citet{zhou2020neural} studied it from a theoretical perspective by providing near-optimal regret bound for neural-network-based UCB and its associated algorithm NeuralUCB. \citet{zhang2020neural} proposed Neural Thompson Sampling, which estimates the variance from the neural tangent features of the corresponding neural network and proves regret bounds matching other contextual bandits algorithms. Different from these works, we address the challenges of deploying bandit algorithms in real-world, large-scale systems, discuss its benefit in exploration, and the lessons learned to bridge the gap between algorithms to applications.
\begin{figure}[!htbp]
    \centering
    \includegraphics[width=\linewidth]{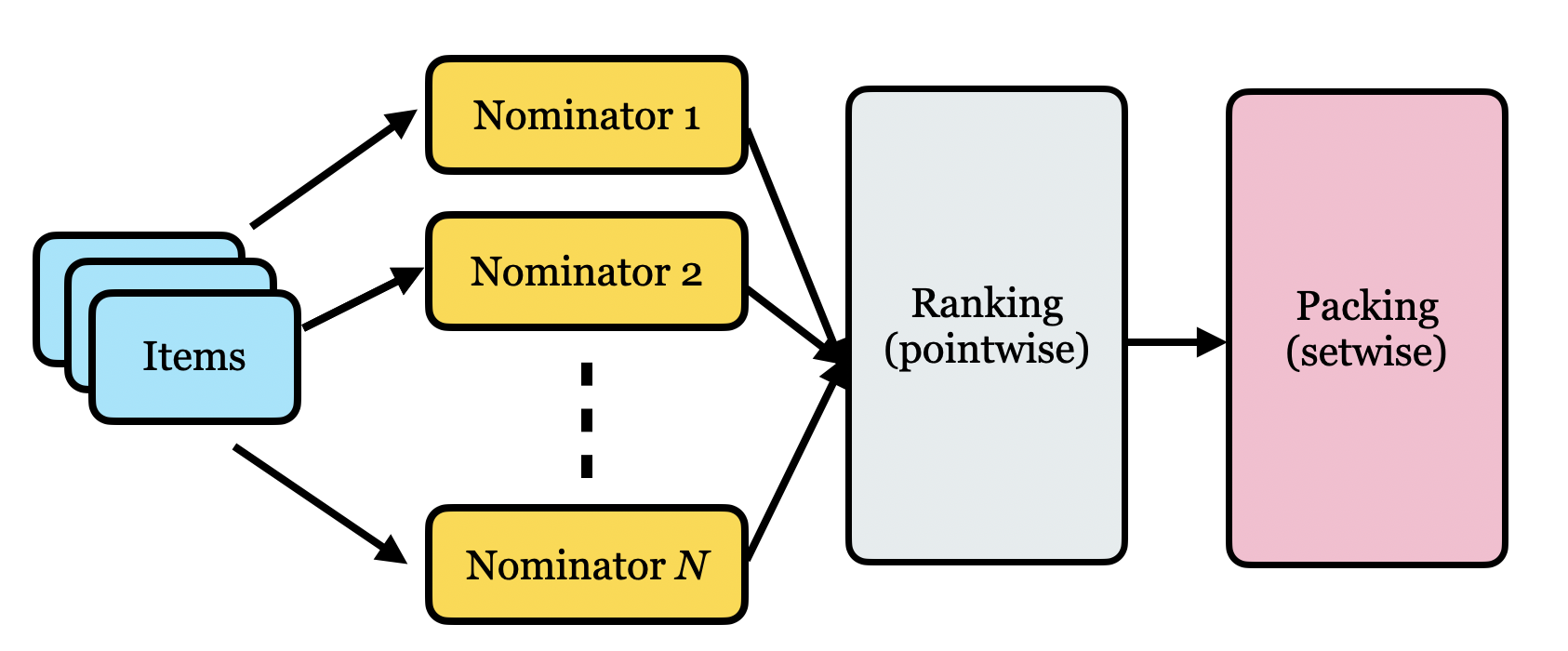}
    \caption{The multi-stage recommender system, with candidate generation in the first stage followed by pointwise ranking and setwise packing.}
    \label{fig:pipeline}
\end{figure}

\section{Background on the overall system}
\label{sec: background}
We study the benefit of exploration and the potential of Neural Linear Bandits for exploration on one of the largest short-form video recommendation platforms serving billions of users, over a period of three months. The production recommender system is a multi-stage system, as illustrate in Figure~\ref{fig:pipeline}. The first stage consists of multiple retrieval systems to identify and nominate promising candidates from the overall content corpus of billions. The second stage involves a ranking system to score and order the pool of candidates (in the order of hundreds), followed by the final packing stage to achieve different business goals and diversify the whole slate.
Unless otherwise specified, we use the production recommender system without the explicit exploration strategy as the control arm in all online A/B testing. We will describe the corresponding treatment arms separately in different experiments.
\section{Long-term Value of Exploration through Enlarged Corpus}
\label{sec: corpus}
In this section, we study the benefit of exploration through corpus change.
In summary, exploration in uncertain regions increases the exposure and discoverability of fresh and tail contents and changes the overall corpus distribution, which in return improves the long-term user experience. We first define the corpus metrics, i.e., discoverable corpus; then introduce a new \emph{user-corpus-codiverted} experiment framework to measure the benefit of exploration on discoverable corpus.
Lastly we present a long-term study showing the effect of corpus change on user experience.

\subsection{Corpus Metrics} 

Exploitation based systems lead to highly skewed corpus distribution, with a small slice of the corpus command a large amount of user interactions, while remaining corpus are hardly surfaced and discovered. We propose a corpus metric that captures the corpus distribution in the number of interactions each video receives. To avoid any confounding factors that might arise from a piece of content having better performance (i.e., receiving more interactions) simply because it receives more impressions from the exploration treatment,
we use \emph{post-exploration} corpus performance to measure the exploration effect. 
Specifically, we set a \emph{graduation} threshold of $X'$ positive user interactions for the contents. In other words, once a piece of content has received more than $X'$ positive user interactions, it is no longer eligible for further exploration treatment. At this point, the content enters the \emph{post-exploration} stage, and needs to survive on its own.
That is to say, the exploration treatment is used to bootstrap cold-start content, but the content's success still mainly depends on its own quality and relevance to the audience once the entry barrier is removed. Given this, we formally define the \emph{Discoverable Corpus $@X, Y$} for a system $\pi$ as:
\begin{align*}
        &\# \text{ of contents receiving more than $X$ positive user interactions } \\  &\text{ within time period $Y$ post-exploration}.
\end{align*}
This metric captures the change in quantity for a range of contents, from tail to head. For small $X$, it measures the performance of the tail content, while for large $X$, it measures the growth of the head and popular content. Ideally, the better the exploration-capability of the system is, the larger the Discoverable Corpus $@X, Y$days for various $X$ buckets, while keeping a relatively neutral user experience as a guardrail\footnote{Our goal is not to solely focus on a pure exploration-based system that maximizes the corpus metrics. Instead, we aim to strike a balance between exploration and exploitation that maximizes the corpus metrics without compromising user engagement too much.}. 
The time window used for evaluation, i.e., $Y$ prescribes the time window allowing newly-explored corpus to grow. In our expriments, we use a 7-day window to capture short-term corpus growth, and 3-month window for long-term growth.


\subsection{User-Corpus-CoDiverted Experiment}
Traditional user-diverted A/B testing~\citep{imbens2015causal, kohavi2020trustworthy} provides a powerful tool to measure the effect of any recommendation changes on the \emph{user} side. In the user-diverted A/B testing, we randomly assign users to the control and treatment group, expose them to the corresponding treatments, and compare the user-side metrics, such as the number of clicks, dwell time, satisfaction survey responses between the two arms.  
These experiments however, cannot capture any corpus change, such as the number of contents receiving more impressions or user interactions because of the exploration treatment. As the two arms share the same corpus, any treatment effect on the corpus will leak between arms. 

\begin{figure}
    \centering
    \includegraphics[width=0.7\linewidth]{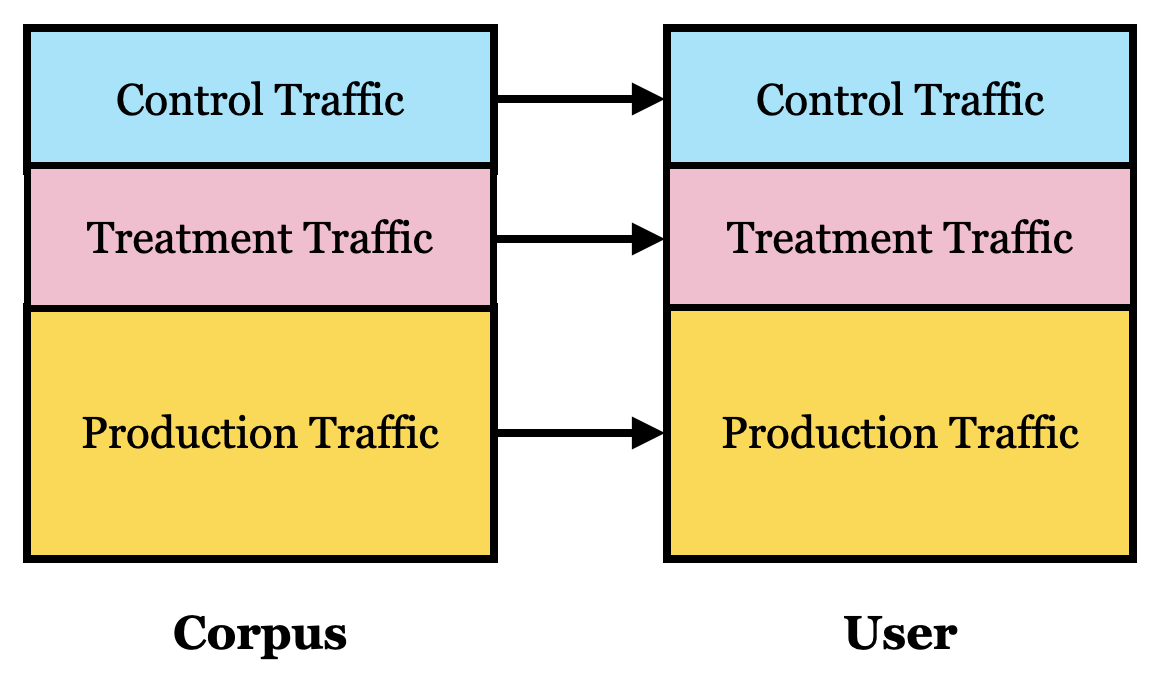}
    \caption{User-Corpus-CoDiverted experiment diagram}
    \label{fig:corpus_diverted}
    \vspace{-0.3cm}
\end{figure}

We thus propose \emph{user-corpus-codiverted} A/B testing, which is an instance of the Multiple Randomization Designs (MRD)~\citep{bajari2021multiple} by designing the specific assignment matrix. Particularly, it randomly assigns x\% of the corpus\footnote{Note that we often further restrict the diversion to be by content providers. In other words, items belonging to the same provider are assigned to the same arm, to avoid treatment leakage between content belonging to the same provider.} into control and treatment arms, in addition to randomly exposing x\% users \emph{in proportion} into control and experiment arms. As shown in Figure~\ref{fig:corpus_diverted}, the users in the control arm only receive recommendations from the control arm's corpus, similarly for the treatment arm.  Compared with the original user-diverted experiment, the random splitting of the corpus prevents treatment leakage and allows measuring the treatment effect on corpus-based metrics. We keep the user and corpus in proportion, e.g., $5\%$ of users explore $5\%$ corpus, so the effectiveness of the exploration treatment is consistent with full deployment when $100\%$ of users are exploring the entire corpus. Otherwise, as one can imagine, using $5\%$ of user traffic to explore the entire corpus ($100\%$) will result in minimal changes to the corpus distribution.

\subsection{Exploration Increases Discoverable Corpus}
\label{subsec: corpus_increase}
\paragraph{Setup:} We conducted a user-corpus-codiverted live experiment where 1) the control arm runs the exploitation-based system as pictured in Figure~\ref{fig:pipeline} to fill all the slots on the platform; 2) the treatment arm runs a simple exploration-based system which reserves dedicated slots for fresh and tail content only\footnote{Fresh \content~ refers to content that is $<X$ days old and tail \content~ refers to the one received less than $Y$ lifetime positive user interactions.}, while filling the other slots using the same system as in the control arm. The exploration system uses a nominator to nominate fresh and tail candidates based on their similarity to historical user consumptions. The nominated candidates are ranked by the same ranking system as in the control arm.

We first examine the \emph{short-term benefit} of the exploration system by measuring the Discoverable Corpus $@X_0, 7 $-day period.
\begin{figure}[!ht]
    \centering
    \includegraphics[width=\linewidth]{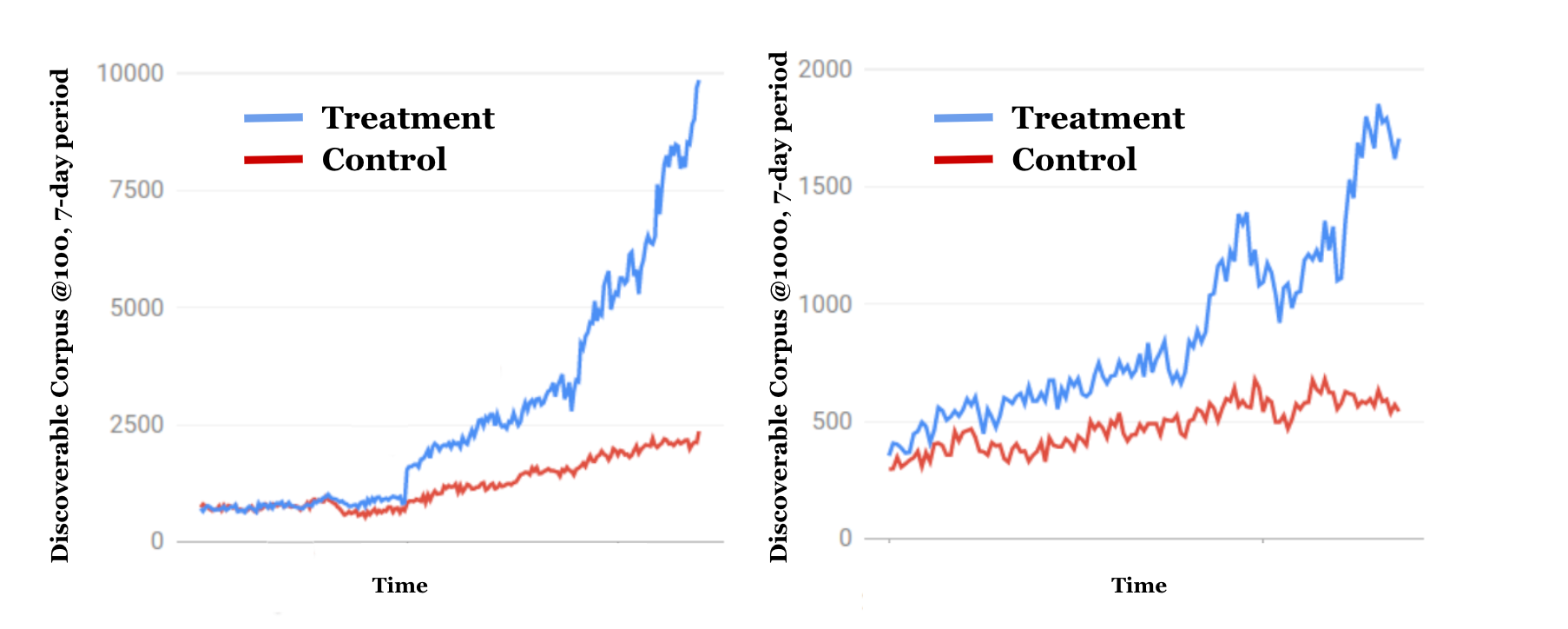}
   \caption{Discoverable Corpus $@100, 7 $-day period (left) and Discoverable Corpus $@1000, 7 $-day period (right) for both control and treatment arms.}
    \label{fig:short_term_vad}
\end{figure}
As shown in Figure~\ref{fig:short_term_vad}, we observe a significantly increased number of Discoverable Corpus $@100, 7 $-day period (left) and Discoverable Corpus $@1000, 7 $-day period (right) (i.e., the number of \contents~ receiving more than $X_0=100$ and $X_0=1000$ post-exploration positive interactions). This validates that the exploration-based system is effective in boosting the number of \contents~ that achieve early success in a short period. Furthermore, the gap between the control and treatment groups continues to widen over time,  which we find is due to content providers in the treatment arm creating more discoverable contents than the ones in the control arm. Although we will not delve into the details of how exploration benefits content providers due to limited space, this topic is worth investigating in future research.

While achieving early success for \content~ in the short term is important, it does not necessarily guarantee sustained long-term growth of the corpus. It is ideal for the exploration system to identify high-quality content that has the potential to go viral after the initial bootstrapping. To assess the \emph{long-term post-exploration growth} of the \contents~,
we analyze the Discoverable Corpus $@X_l, 3 $-month period ($X_l, X_l \gg X_0$) for different $X_l$ buckets. 
As shown in Table~\ref{table:mid_term}, the exploration treatment consistently increases the Discoverable Corpus $@X_l, 3 $-month period across different $X_l$ buckets. It is noteworthy that the percentage of increment remains remarkably consistent across the different $X_l$ buckets, roughly around 50\%. 
\begin{table}[htp]
\caption{The change in Discoverable Corpus $@X_l, 3 $-month period between control and treatment arms.}
\begin{tabular}{ cccccc }
\hline
$X_l$ & 100 & 1000 & 10K & 1M & 10M \\
 & (in 7 days) & (in 7 days) & & & \\
\hline
Change & $+119.4\%$ & $+58.5\%$ & $+48.2\%$&$+51.0\%$ & $+53.8\%$\\
\hline
\end{tabular}
\label{table:mid_term}
\end{table}

\begin{figure}[!ht]
    \centering
    \includegraphics[width=\linewidth]{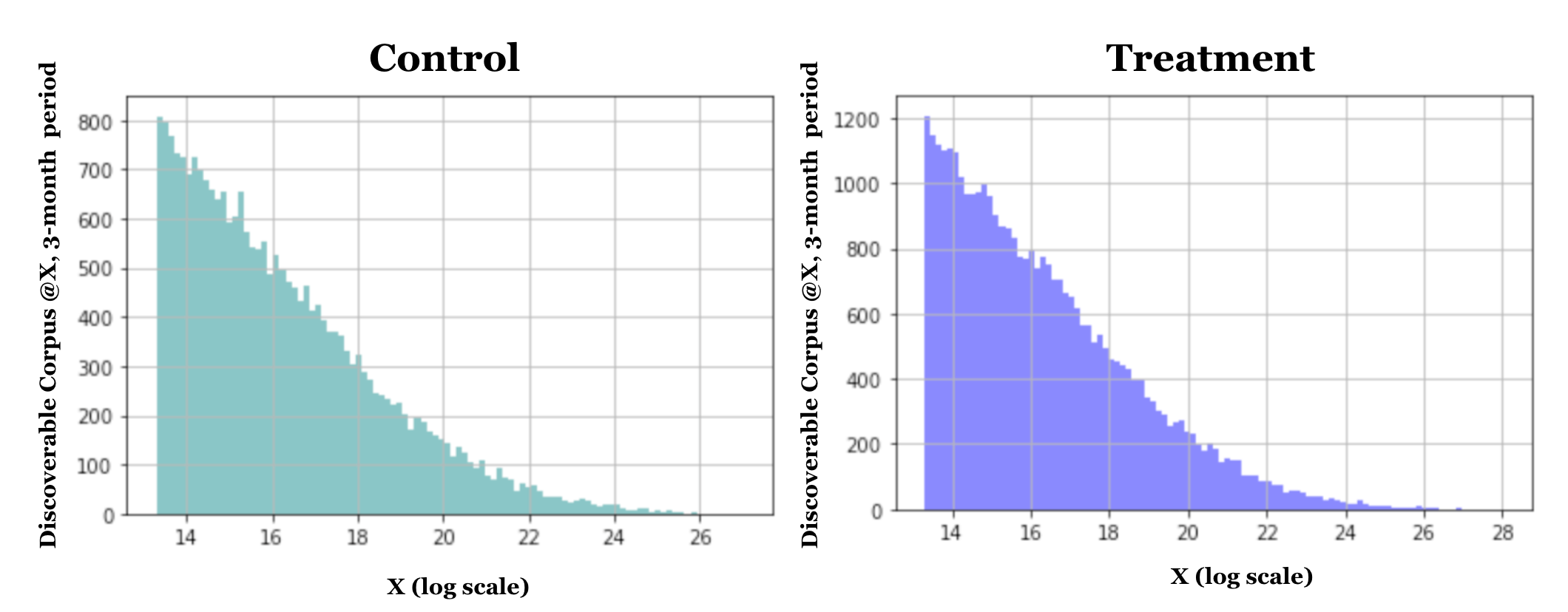}
    \caption{The histogram of the Discoverable Corpus $@X, 3 $-month period (number of \contents~ receiving at least $X$ positive feedback) for various $X$ values, between control and treatment arms. The $x$-axis shows the value of $X$ (in log scale), and the y-axis denotes the Discoverable Corpus $@X, 3 $-month period.}
    \label{fig:mid_term_hist}
\vspace{-0.2in}
\end{figure}

To examine the full spectrum of the \content~ quality distribution, we use the Discoverable Corpus $@X, 3 $-month period at various threshold $X$ as the quality indicator.  Figure~\ref{fig:mid_term_hist} shows the histograms of Discoverable Corpus $@X, 3 $-month period
for both control (left) and treatment (right) arms. We focused on \contents~ with at least 10k post-exploration positive feedback (i.e., $X\geq 10K$) to zoom in on the "high-quality" region of the corpus. The x-axis shows the number of post-exploration positive user interactions on a logarithmic scale (i.e., $X$), and the y-axis plots the number of contents in the specific bucket (i.e., Discoverable Corpus $@X, 3 $-month period).
While the numbers in the treatment arm (right) are higher, we observe a remarkably similar distribution between the two, implying that the quality distribution of \contents~ discovered through the exploration system is comparable to that of the original \contents~. In other words, exploration not only helps \contents~ achieve initial success, but also discovers the high-quality ones which will eventually reach viral. 

\subsection{Long Term Value of Enlarged Discoverable Corpus}
\label{subsec: corpus_long_term}
The discussion above established that exploration enlarges discoverable corpus proportionally in different interaction buckets, and discovers both "future" head and tail \contents~. In this section, we close the loop of the argument by connecting the change in the discoverable corpus size to the long-term user experience. To quantify the user satisfaction, we use a metric that counts the number of daily active users with satisfactory interactions (based on satisfaction survey prediction) on the platform, which we refer to as \emph{satisfied daily active users} throughout the paper.

The crux of the study is to allow each user to access a reduced and fixed \footnote{To ensure that each user is only allowed access to a \emph{fixed} reduced corpus, we use the same seed $s_u$ for different requests from the same user (see Algorithm~\ref{alg:corpus_ablation_procedure}). However, to prevent any specific content from being dropped for all user traffic, we choose different seeds across users.} corpus $\mathcal{C}' \subset \mathcal{C}$ by removing any nominations $c\in \mathcal{C} \setminus \mathcal{C'}$ and observe the change in the number of satisfied daily active users~\footnote{To eliminate any confounding factors, i.e., ensuring the same number of candidates are scored in the second stage (ranking) after filtering nomination candidates by the reduced corpus, we increased the number of nominations in the first stage accordingly.}. 
We conduct the ablation study for 4 weeks, with both the control and treatment arms running the same multi-stage recommender system as depicted in Figure~\ref{fig:pipeline}. Each arm takes $5\%$ of the overall traffic.
The control arm receives all candidates outputted by the nominators, while the treatment randomly filters $x\%$ of the corpus from the platform\footnote{There are various ways to perturb the content corpus and observe changes in user experience. One alternative is to directly removing various percentages of the ``exploratory" content, however, this approach might be subject to the choice of the exploration system used in the filtering. Here, we present this more general approach in random filtering to decrease the discoverable corpus size.}, using the Corpus Ablation Procedure detailed in Appendix~\ref{app:corpus_ablation} Algorithm~\ref{alg:corpus_ablation_procedure}.

\begin{figure}[!ht]
    \centering
    \includegraphics[width=1.0\linewidth]{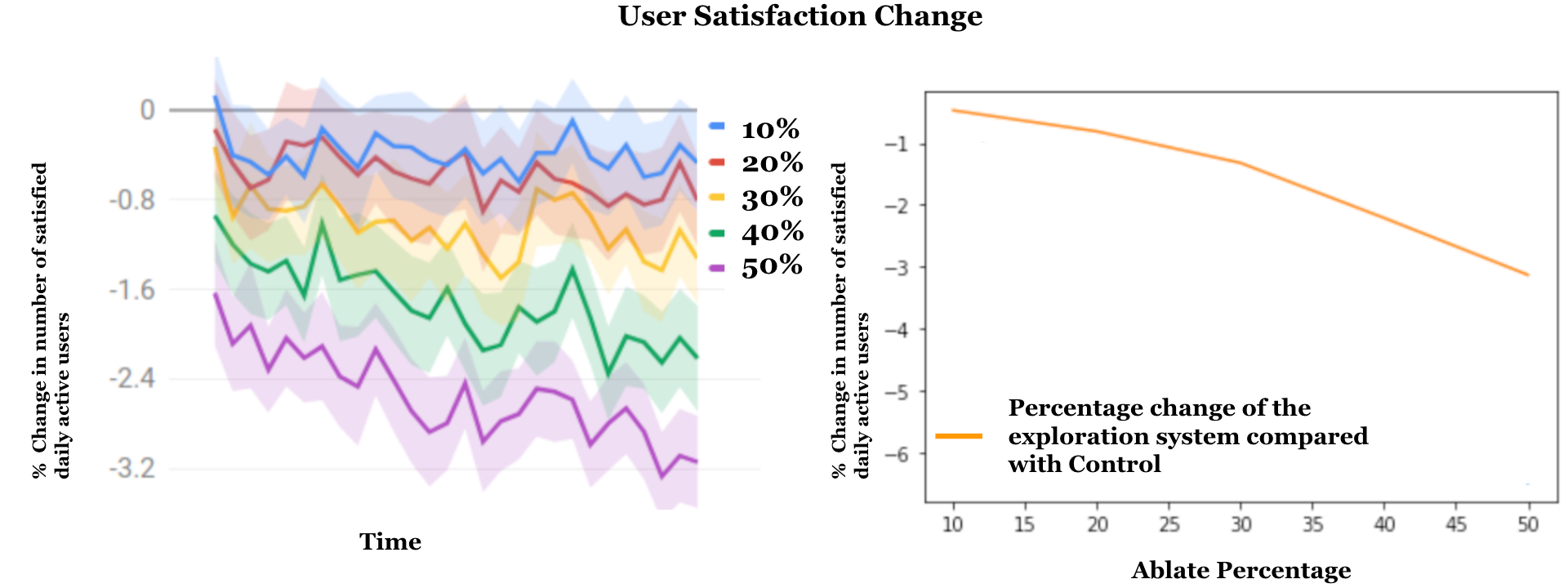}
    \caption{Left: the percentage change of the number of satisfied daily active users across time, for different ablation size $x\%$.  Right: Linear interpolation of the change in terms of number of satisfied daily active users w.r.t. discoverable corpus size change. }
    \label{fig:corpus_size_exp}
\end{figure}

The results of the study are presented in Figure~\ref{fig:corpus_size_exp}. From the left figure, we observe a significant decrease in the satisfied daily active users across different ablation sizes. Moreover, the negative impact of the ablation grows over time, suggesting that it has a long-lasting negative effect on long-term user satisfaction. Interestingly, the right figure shows a monotone relationship between discoverable corpus size change and the number of satisfied daily active users (roughly linear), from which we hypothesize increasing discoverable corpus size will lead to positive user experience.
However, it is worth mentioning that the linear relationship might only hold for a specific range of corpus size. Additionally, growing the corpus could have a saturating effect when the discoverable corpus reaches a certain size. Determining the exact nature of this relationship is an exciting direction for future research, but beyond the scope of this paper.
In summary, exploration increases the size of discoverable corpus, which further translate into long term user satisfaction gain.

\section{Neural Linear Bandit Based Ranking System}
Given the established long-term value of exploration, we further investigate efficient exploration techniques that can operate in real-world, large-scale industrial systems.
To this end, we examine the potential of Neural Linear Bandit (\shortname) for ranking, which builds upon the NeuralLinear algorithm~\citep{riquelme2018deep}. We choose NeuralLinear as the backbone algorithm as it leverages the representation power of DNNs, which is the foundation for many industrial systems, and computes the variance efficiently.
We focus our discussion on the implementations and modifications we made to incorporate it into the current industrial system pipeline, as well as the challenges and exciting future directions for building exploration-based recommender systems.

\subsection{Neural Linear Bandit}
We frame recommendation as a contextual bandit problem:
given the overall time horizon $T$, at each time-step $t\in[T]$, a user comes to the platform with contextual feature $\ub_t \in \RR^{d_u}$. The system $\pi$ reacts to the user by selecting and presenting the action (i.e., content) $a_t \in \Acal$ with feature $\ab_t\in \RR^{d_a}$. Then the system receives the feedback/reward (such as click, completion rate, likes, etc) $\rb_t \sim P(\cdot|\ub_t, \ab_t)$ with $\EE[\rb_t]=f(\ub_t, \ab_t)$ and $f$ being some unknown relevance function. The system then updates its model using the newly collected interaction data. The goal of the system $\pi$ is to minimize the following regret:
\begin{equation}
\label{eqn:regret}
    R_T(\pi) := \EE_{\mathbf{u}_t\sim P_\mathcal{U},  a_t\sim \pi(\cdot|\ub_t), \rb_t\sim P(\cdot|\ub_t, \ab_t) }\bigg[\sum_{t=1}^T \rb(\ub_t, \ab^*_t) - \rb(\ub_t, \ab_t)\bigg]
\end{equation}
with $\ab^*_t:=\arg\max_{a\in \Acal} \rb(\ub_t, a)$ denotes the optimal action in hindsight, and $\ab_t$ is the  action selected from system $\pi$. 
Popular algorithms such as UCB~\citep{chu2011contextual, li2010contextual} and Thompson Sampling~\citep{agrawal2013thompson, zong2016cascading,cheung2019thompson, kveton2022value} with linear payoff function~\cite{chu2011contextual, agrawal2013thompson} have been studied extensively for its closed form update. 
Linear models however are restrictive in representation power. 
We adopt NeuralLinear~\citep{riquelme2018deep}, a variant of Thompson Sampling algorithm that performs the linear Bayesian regression on the top of the last layer features of the neural network, which nicely combines the representation power of the deep neural network and the simplicity of the uncertainty estimate from linear models. The underlying assumption is given by:
\begin{assum}
\label{assum:linear_last_layer}
There exists a representation function $\phi:\RR^{d_u}\times \RR^{d_a}\to \RR^d$ and an unknown parameter $\beta\in\RR^d$, such that for all user-content pair $(\ub, \ab)$, the mean reward $\rb(\ub, \ab)$ is linear in the representation $\phi(\ub, \ab)$ with
\begin{equation}
    \EE[\rb] = \phi(\ub, \ab)^T\beta
\end{equation}
\end{assum}
Modern deep learning-powered recommendation models often take a (generalized) linear regression layer from the last layer embedding to the target. We thus use the learned last layer embedding as the representation function $\phi$ and empirically verify that Assumption~\ref{assum:linear_last_layer} is roughly held by examining various evaluation metrics.
It then assumes the reward $\rb_t$, with embedding $\phi(\ub_t,\ab_t)$ and parameter $\beta$, follows the Gaussian distribution $\Ncal(\phi(\ub_t, \ab_t)^T\beta, \sigma^2)$ with noise $\sigma^2$. At time step $t$, \shortname~ samples action $\ab_t$ under the new user context $\ub_t$ and observes reward $\rb_t$. Assuming the prior distribution for the unknown parameter $\beta$ is given by $\Ncal(\hat{\beta}_{t-1}, \sigma^2 \Sigma^{-1}_{t-1})$, it then updates the posterior distribution for $\beta$ as:

\begin{equation}
    \PP(\beta|\rb_t) \propto \PP(\beta)\PP(\rb_t|\beta) \propto \Ncal(\hat{\beta}_{t}, \sigma^2 \Sigma^{-1}_{t})
\end{equation}
The covariance matrix $\Sigma_t$ and parameter estimate $\hat{\beta}_t$ are defined as
\begin{align}
\label{eqn: update}
&\Sigma_t := \epsilon I_d + \sum_{\tau=1}^{t} \phi(\ub_\tau, \ab_\tau) \phi(\ub_\tau, \ab_\tau)^T , \hat{\beta}_t := \Sigma^{-1}_t \big(\sum_{\tau=1}^{t}\phi(\ub_\tau, \ab_\tau) \rb_\tau \big)
\end{align}
with $\epsilon$ is the hyper-parameter that controls the regularization strength. The posterior distribution of the reward $\rb(\ub, \ab)$ at round $t+1$ is therefore given by 
\begin{equation}
\label{eqn:post_dist}
    \Ncal\big(\phi(\ub, \ab)^T\hat{\beta}_t, \sigma^{2}\phi(\ub, \ab)^T \Sigma^{-1}_t \phi(\ub, \ab)\big)
\end{equation}
Neural Linear Bandit then samples the reward for each action from this posterior distribution, and pulls the action which attains the highest reward.

\subsection{Implementation}
\label{subsec:implementation}
\begin{figure}[!htbp]
    \centering
    \includegraphics[width=0.8\linewidth]{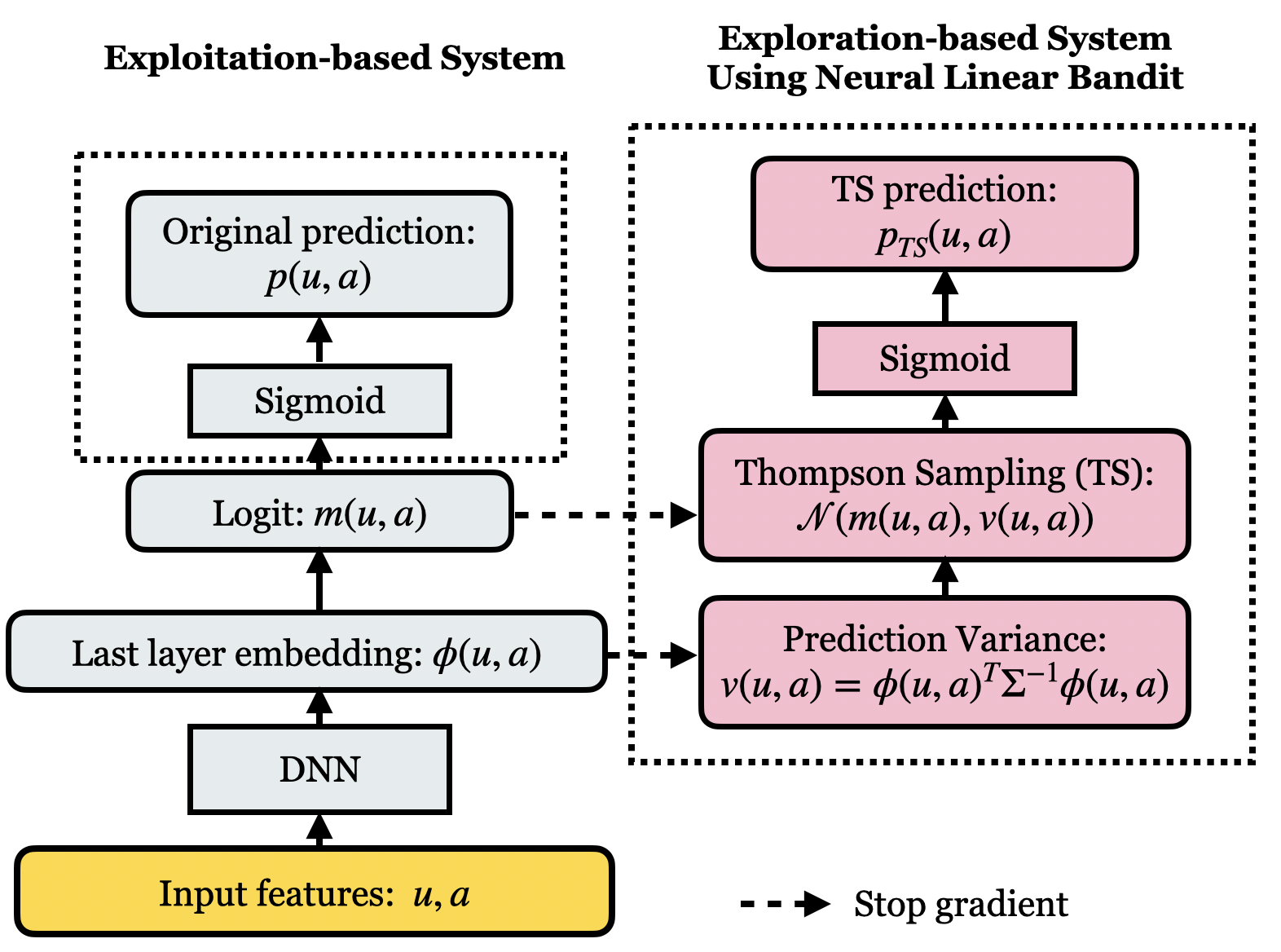}
    \caption{The model architecture for the exploitation-based system (control) and the exploration-based system using Neural Linear Bandit (treatment), on a classification task.
    }
    \label{fig:model_arch}
\end{figure}

Despite its simplicity, fitting \shortname~ directly into the industrial training and serving pipeline however, faces challenges due to 
(1). the high frequency of model updates; 
(2). the stable and efficient implementation of the variance estimate (i.e., $\sigma^{2}\phi(\ub, \ab)^T \Sigma^{-1}_t \phi(\ub, \ab)$) as well as 
(3). the extension of the model beyond regression tasks, as some labels/rewards are binary, such as clicks, likes, etc. 
In this section, we discuss practical implementations and different design choices we made to enable the use of \shortname~ in the industrial recommender system training pipeline. We provide the detailed algorithm shown in Algorithm~\ref{alg:overall_alg} with overall architecture to build the Neural Linear Bandits into an industrial ranking system in Figure~\ref{fig:model_arch}, and empirically evaluate it in Section~\ref{sec: experiment}. 
\paragraph{Setup.}
We apply the Neural Linear Bandit to the ranking stage of the system (Figure~\ref{fig:pipeline}), which ranks hundreds to thousands of candidates nominated from the first candidate generation stage. 
To use \shortname, we replace the original prediction score of a candidate with a sample from the posterior distribution of the prediction, and then rank all the candidates greedily in decreasing order. We choose one particular classification task for simplicity, i.e., predicting whether or not a user completes\footnote{We mark the content as complete if the user consumes more than $x\%$ of the contents.} the recommended content
as the binary reward $\rb$. The underlying model for this classification task is a deep neural network architecture shown in Figure~\ref{fig:model_arch}. The user features $\ub$ and content features $\ab$ are concatenated as input features and pass through several dense layers to predict the final score of completion. We use the 128-dimensional last layer embedding of the neural network $\phi(\ub, \ab)$\footnote{In experiment, we also concatenate the last layer embedding with several user and content features for better predictability of the relevance and uncertainty.} as the contextual features used in the Neural Linear Bandit, as shown in the pink blocks. 

\paragraph{Investigation 1: Model Update Frequency.} The \shortname~ algorithm updates the covariance matrix $\Sigma_t$ per sample. It then computes the prediction variance which requires computationally intensive inversion calculation ($\Sigma^{-1}_t$) at every sample as in equation~\ref{eqn: update} and ~\ref{eqn:post_dist}. 
On the other hand, most industrial recommendation models are trained in a batch setup~\citep{jeunen2021top, bendada2020carousel}, where the model is continuously trained on a large set of logs, check-pointed, and exported daily or multiple times a day. To be more efficient, we decompose the updates of \shortname~ into two stages:
(1). In the \emph{training} stage, we keep updating the covariance matrix $\Sigma_t$ using Equation~\ref{eqn: update} as we process each training data in the log, with the (changing) learned embedding function $\phi$; When the training finishes, we update the precision matrix $\Sigma^{-1}_t$ once, which ensures that the expensive inversion calculation is performed in a much less frequent manner, i.e., only once per training run.
(2). In the \emph{serving} update, for each user $\ub$ comes to the system, we calculate the prediction variance using the fixed precision matrix, and draws the predicted reward from the posterior distribution\footnote{As discussed in the later, for the mean estimate $\phi(\ub, \ab)^T\beta$ used in the posterior distribution, we directly use the predictions derived from the original system.} defined in Equation~\ref{eqn:post_dist}. We present the top $K$ results to the user and collect corresponding feedback.

\paragraph{Investigation 2: Stability and Variants of Uncertainty Calculation.} 
The uncertainty estimate requires the inversion of the covariance matrix $\Sigma^{-1}$, which brings the stability issue when the covariance matrix is close to singular.
We investigate two stable and efficient ways for uncertainty calculation: Pseudo-inverse~\citep{penrose1955generalized} and Cholesky decomposition~\citep{press2007numerical, krishnamoorthy2013matrix}, by comparing their computational cost and estimation accuracy. 
\emph{Pseudo-inverse} provides a natural way to solve the under-determined linear systems. To avoid singularity in the precision matrix update, we replace the inverse $\Sigma^{-1}$ with its pseudo-inverse variant $\Sigma^{\dagger}$, which is equivalent to $\Sigma^{-1}$ when the covariance matrix is of full rank. It provides the unique solution with the minimum $L_2$ norm on the parameter $\beta$ when the system is under-determined.
\emph{Cholesky decomposition} is another popular method to calculate the parameter and the quadratic uncertainty term, avoiding explicit matrix inversion. Specifically, it calculates the Cholesky decomposition of the positive definite covariance matrix $\Sigma = LL^T$, with $L$ being the lower-triangular matrix. Following this, the variance term in Equation~\ref{eqn:post_dist} could be rewritten as:
    \begin{align}
        \sigma^{2}\phi(\ub, \ab)^T \Sigma^{-1} \phi(\ub, \ab) &= \sigma^{2}\phi(\ub, \ab)^T (LL^T)^{-1} \phi(\ub, \ab) \\
        &=\sigma^{2}\phi(\ub, \ab)^T L^{-T}L^{-1} \phi(\ub, \ab) \\
        &=\sigma^{2}z(\ub, \ab)^T z(\ub, \ab)
    \end{align}
    with $z(\ub, \ab):=L^{-1} \phi(\ub, \ab)$ and could be easily solved using the forward substitution $Lz(\ub, \ab) = \phi(\ub, \ab)$ with only $\Ocal(d^2)$ complexity. Similarly, given the Cholesky decomposition, the parameter $\beta$ could be solved by continuously solving two linear systems with the lower-triangular matrix $L$. 

\begin{figure}[!ht]
    \centering
    \includegraphics[width=0.45\linewidth]{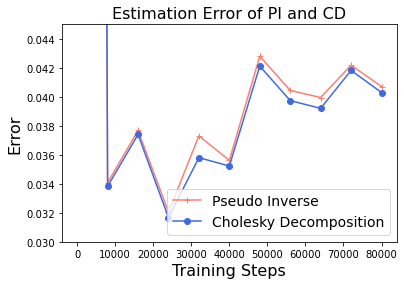}
    \includegraphics[width=0.45\linewidth]{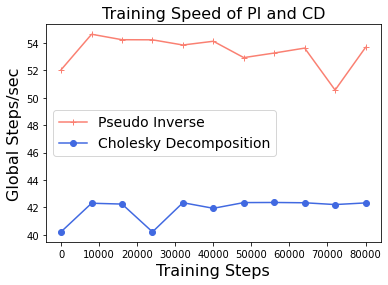}
    \caption{Left: The estimation error (compared with neural network prediction) of Pseudo-inverse vs. Cholesky decomposition, the lower the better. Right: The training speed, the higher the better. }
    \label{fig:cholesky_vs_inverse}
    \vspace{-0.1cm}
\end{figure}

Here we investigate the pros and cons of these two approaches regarding estimation accuracy and training speed. Figure~\ref{fig:cholesky_vs_inverse} (left) plots the mean absolute difference between the predicted reward 
and the neural network prediction $\hat{r}(\ub, \ab)$, i.e., $|\phi(\ub, \ab)^T\hat{\beta}-\hat{r}(\ub, \ab)|$, with $\hat{r}(\ub, \ab)$ served as a surrogate to the ground-truth $\EE[r(\ub,\ab)]$. We see that Cholesky decomposition does give a smaller error, which shows its advantage in the accuracy and stability of the solution. The training speed is shown in Figure~\ref{fig:cholesky_vs_inverse} (right), and we do see the pseudo-inverse is much faster compared with Cholesky decomposition, as in our case, the size of the matrix is small with $d=128$. 
Hence we go forward with the Pseudo-inverse option. 
\begin{algorithm}[!htbp]
\SetAlgoLined
\textbf{Input:} Total number of training runs $T$; Total number of batches per training run $H$; Initialize the model $f$ with parameter $\theta_0$ and last layer representation $\phi_0$; Initialize the dataset $\Dcal_0 = \emptyset$; Initialize Neural Linear Bandit parameter $\Sigma_{1,0}=\Sigma_0:=\epsilon\mathbf{I}$, $\beta_0=\textbf{0}$; noise parameter $\sigma^2$\\
\For{each training run $t=1,2,\cdots$}{
\textbf{Training Stage:}\\
Collect user log $\Dcal_{t-1}$;\\
\For{each batch $h=1,2,\cdots, H$ with batch data $\Dcal_{t-1,h}$}{
Update the covariance matrix:  $$\Sigma_{t,h}:=\Sigma_{t,h-1}+\sum_{(\ub,\ab) \in \Dcal_{t-1,h}}\phi_{\theta_{t,h-1}}(\ub,\ab)\phi_{\theta_{t,h-1}}(\ub,\ab)^T$$\\
Update the parameter of the ranking model $\theta_{t,h-1}\to\theta_{t,h}$, its associated last layer parameter $\phi_{\theta_{t,h-1}}\to\phi_{\theta_{t,h}}$ and the parameter $\beta_{t, h-1}\to\beta_{t, h}$ using stochastic gradient descent on batch data $\Dcal_{t-1, h}$.
}
Update the precision matrix $\Sigma^{\dagger}_{t}$ as pseudo-inverse of $\Sigma_{t,H}$; \\
Take $\Sigma_{t+1,0}:= \Sigma_{t,H}$, $\theta_t:=\theta_{t,H}$ and $\beta_t:=\beta_{t,H}$, and push the model $\theta_t, \beta_t, \Sigma^{\dagger}_{t}$ for serving.\\
\textbf{Serving Stage:}\\
\For{each user $\ub$ comes to the system, for all nominations (actions) $\ab \in\Acal$ }{
Calculate the mean: $m_{\theta_t}(\ub, \ab):=\phi_{\theta_t}(\ub,\ab)^T \beta_t$, and the variance of the posterior distribution:
$$v_t(\ub,\ab):=\sigma^2\phi_{\theta_t}^T(\ub,\ab)\Sigma^{\dagger}_t\phi_{\theta_t}(\ub,\ab)$$\\
Sample $m_{TS}(\ub,\ab)$ from posterior distribution $\Ncal(m_{\theta_t}(\ub,\ab), v_t(\ub,\ab))$\\
Modify the original ranking score with $p_{TS}(\ub,\ab):=\mu(m_{TS}(\ub,\ab))$, and $\mu$ being the sigmoid function;\\
Present the top $K$ ranked contents $\Acal_k$ to the user;\\
Collect user feedback $\rb(\ub,\ab)$ for $\ab\in\Acal_k$;\\
Update log $\Dcal_{t}\to\Dcal_{t-1}\bigcup\{(\ub,\ab,\rb(\ub,\ab))\}_{\ab\in\Acal_k}$
}
}
\vspace{-0.1cm}
\caption{Neural Linear Bandit (Classification Task)}
\label{alg:overall_alg}
\end{algorithm}

\paragraph{Investigation 3: Extension to Classification Task.} 
In cases such as predicting completion, clicks and likes, the task is classification rather than regression. It is well known that generalized linear models (for example, logistic regression) show stronger performance than linear models when the reward is binary~\citep{filippi2010parametric}. Prior work has studied efficient exploration techniques when the payoff is a generalized linear model of the contextual feature, i.e., $\EE[\rb] = \mu(\phi(\ub, \ab)^T\beta)$ with known link function $\mu$ and unknown parameter $\beta$, for example, GLM-UCB~\citep{li2017provably} and GLM-TSL~\citep{kveton2020randomized}. In the generalized linear models (GLM) setting, the challenge lies in the fact that the maximum likelihood estimation (MLE) of $\beta$ no longer admits a one-off closed-form update as in linear models~\footnote{In linear models, one can derive $\beta_t$ in a one-off close form update as in eq~(\ref{eqn: update}) by aggregating the reward weighted context features (the term in the parenthesis) per-sample.}, i.e., in GLM, we need to get the MLE for $\beta$ at every time step $t$ by solving~\citep{li2017provably, kveton2020randomized}:
\begin{equation}
    \sum_{\tau=1}^{t-1} (\rb_{\tau} - \mu(\phi(\ub_\tau,\ab_\tau)^T\beta))\phi(\ub_\tau,\ab_\tau) = 0
\end{equation}
which uses all the previous observations at each round and brings expensive per-sample gradient updates.
However, it is easy to see that the $\hat{\beta}$ is only needed to predict the posterior distribution mean of the reward $\mu(\phi(\ub,\ab)^T\hat{\beta})$.
A cheap surrogate exists for it, i.e., the \emph{logit} of the original binary label prediction $\hat{\rb}(\ub,\ab)$, which is a by-product of the current system and provides a consistent estimate of the mean. 
When choosing the optimal action to play, we select the action $\ab$ that maximizes $\phi(\ub,\ab)^T\beta$ as since it is equivalent to the $\text{argmax }\mu(\phi(\ub,\ab)^T\beta)$ when $\mu$ is the stictly increasing function.
This idea shares a similar flavor with the SGD-TS algorithm proposed by ~\citet{ding2021efficient}, which shows that under diversity assumption on the contextual features, online SGD with TS exploration could achieve $\Tilde{O}(\sqrt{T})$ regret for finite-arm GLM problems. Unlike SGD-TS, we calculate the exact matrix pseudo-inverse for a more accurate uncertainty estimate, instead of approximating it through a diagonal matrix. The uncertainty estimate is the same as the linear case, and we could calculate it by simply maintaining the covariance matrix. After the posterior sampling is done in the linear logit space, we could transform the sample into the original space through the link function $\mu$.

\section{Experiments}
\label{sec: experiment}
We conduct a series of online A/B testing to evaluate the performance of the Neural Linear Bandit based ranking system, on one of the largest short-form video recommendation platforms. We also examine the properties and the reliability of the uncertainty measurement.

We first ran the user-diverted A/B testing with $0.3\%$ traffic on both the control and treatment arms for six weeks to observe user metrics. 
The control arm is the original ranking model in production, and the treatment is the exploration-based ranking system that utilizes the Neural Linear Bandit. For NLB, as discussed in Section~\ref{subsec:implementation}, we update the covariance matrix in streaming fashion while the precision matrix $\Sigma^{\dagger}$ is updated offline per training run, to be consistent with the training pipeline. To ensure stability in the matrix inversion, we set the regularization hyper-parameter with $\epsilon=1e-6$ (Equation~\ref{eqn: update}). To pick the noise parameter $\sigma^2$, we calculate the uncertainty from the ensemble of 5 different training models (which serves as a costly ground-truth measure), and pick the constant hyper-parameter $\sigma^2=10$ such that the uncertainty derived from ensembles and Neural Linear Bandit are roughly in the same order.

\begin{figure}[!ht]
    \centering
    \includegraphics[width=0.8\linewidth]{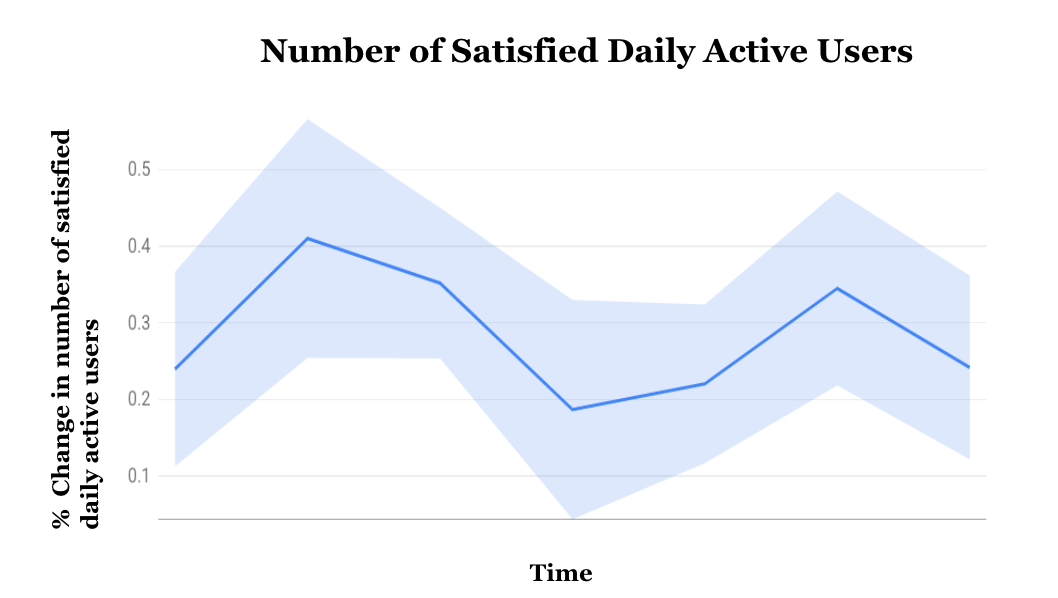}
    \caption{The gain (with the $95\%$ confidence interval) for the Neural Linear Bandit based ranking system in terms of user satisfaction, compared with the the control production system, over a 6-week period.}
    \label{fig:engagement_metric}
    \vspace{-0.2cm}
\end{figure}

\paragraph{How does the Neural Linear Bandit perform, in terms of content freshness and user satisfaction?} 
Intuitively, the uncertainty-based exploration system (e.g., NLB) gives more exposure to fresh and tail content, which changes the overall content corpus distribution and acquires valuable learning signals from these regions, and this further translates to user-engagement gain. 
Table~\ref{tab:fresh_metric} reports the increase in positive interactions on fresh contents published within different time periods.
The time buckets (for example, 1h) in the header row group the \contents~ based on different freshness levels.
The significant increase of positive interactions on \contents~ across different freshness levels verifies that exploration could help the system to explore fresh contents effectively and to acquire valuable learning signals. Interestingly, we also see a stable increase in the number of satisfied daily active users over time, as shown in Figure~\ref{fig:engagement_metric}. We conjecture the gain might come from the following two aspects. First, the system helps users discover novel interests, as we also see a $+1.25\%$ gain in the number of unique topics that a user provides positive interactions on. Meanwhile, users prefer fresh contents on the specific surface specializing in short-form \contents~. 
\begin{table}[!ht]
\centering
\caption{The gain in various freshness related metric.}
\vspace{-0.2cm}
\begin{tabular}{c|c|c}
\hline
Fresh pos. feedback gain & Mean & 95\% CI  \\
\hline
\textbf{1h} & 1.49\% & [1.20, 1.77]\% \\
\textbf{3h} & 1.51\% & [1.24, 1.77]\% \\
\textbf{12h} & 1.45\% & [1.19, 1.71]\% \\
\textbf{1d} & 1.43\% & [1.18, 1.69]\% \\
\textbf{3d} & 2.55\% & [2.30, 2.81]\% \\
\textbf{12d} & 1.16\% & [0.92, 1.41]\% \\
\hline
\end{tabular}
\label{tab:fresh_metric}
\vspace{-0.1in}
\end{table}

\paragraph{What are the properties of the uncertainty estimates and how reliable they are?} 
One of the key components in the Neural Linear Bandit is the quadratic uncertainty term, which captures the strength of the exploration term for different $(\ub, \ab)$ pairs. Though theoretically quantifiable, it remains an interesting question to visualize how the uncertainty differs for different users and content types. To examine this, we select three representative features, with two capturing content properties: (1). the number of days since the contents' publish time (i.e., content age); (2). the number of lifetime positive interactions (i.e., content popularity) and one capturing user property: (3). the total number of interactions a user provided on the platform (i.e., user activity).
We measure the relationship between these features and the uncertainty term using Spearman's rank correlation coefficient, which assesses the monotone relationship between two variables. Table~\ref{table:correlation} reports the Spearman's rank correlation coefficient between the selected three features and the uncertainty calculated by Neural Linear Bandit. Interestingly, one can observe that the current system is more uncertain for fresh and less popular contents while being more or less indifferent to users of different activity levels. Furthermore, we compute the Spearman's rank correlations between the features and the uncertainties obtained from ensemble models, which shows $-0.3$ for both the content features and around $0$ for user features. These results are similar to the ones calculated from the Neural Linear Bandit, suggesting the reliability of the uncertainty estimate.
\vspace{-0.1cm}
\begin{table}[!ht]
\caption{The Spearman's rank correlation (with standard errors over 20 runs) of different features and the uncertainty calculated by Neural Linear Bandit.}
\begin{tabular}{ ccc }
\hline
 & Neural Linear Bandit \\ 
 \hline
Content Age & -0.35 $\pm$ 0.003 \\
Content Popularity & -0.26 $\pm$ 0.003\\
User Activity & 0.02 $\pm$ 0.008 \\
\hline
\end{tabular}
\label{table:correlation}
\end{table}

\paragraph{How does the Neural Linear Bandit affect corpus metric change?} 
To examine the exploration capability of the Neural Linear Bandit, i.e., how it benefits the size of the content corpus, we perform the user-corpus-codiverted experiments with $5\%$ corpus and users diverted to the control and treatment arm separately. For the exploration-based ranking system built upon Neural Linear Bandit, we see a $+5.33\%$ increase for the Discoverable Corpus $@100, 7 $-day period, and another $+5.66\%$ improvement for the Discoverable Corpus $@1000, 7$-day period. 
Compared with the exploitation-based system, Neural Linear Bandit distributes the \contents~ more equitably. Specifically, the improvement in the post-exploration metrics suggests more discoverability of tail \contents~.

\section{Discussion and Future Works}
In this paper, we conduct a systematic investigation into the long-term value of exploration on users through a crucial intermediary in the system, namely the content corpus. We address the measurement challenges, design new metrics and experimental frameworks to capture the benefits of exploration on the content corpus, and establish a connection between the growth of the discoverable corpus and the gain in long-term user satisfaction. We verify them via extensive real-world live experiments in a large-scale commercial recommendation platform and present our valuable findings. 
We further examine Neural Linear Bandit algorithm for building the uncertainty-based exploration system in production. It is worth pointing out that the current setup is tailored for single-task prediction and exploration. In contrast, most modern recommender systems aim to capture multiple rich sources of feedback, and often use multi-task learning in their practical applications.
How to efficiently explore under these more complex multi-task settings is an interesting future direction.

\bibliographystyle{ACM-Reference-Format}
\bibliography{sample-base}


\begin{thebibliography}{38}


\ifx \showCODEN    \undefined \def \showCODEN     #1{\unskip}     \fi
\ifx \showDOI      \undefined \def \showDOI       #1{#1}\fi
\ifx \showISBNx    \undefined \def \showISBNx     #1{\unskip}     \fi
\ifx \showISBNxiii \undefined \def \showISBNxiii  #1{\unskip}     \fi
\ifx \showISSN     \undefined \def \showISSN      #1{\unskip}     \fi
\ifx \showLCCN     \undefined \def \showLCCN      #1{\unskip}     \fi
\ifx \shownote     \undefined \def \shownote      #1{#1}          \fi
\ifx \showarticletitle \undefined \def \showarticletitle #1{#1}   \fi
\ifx \showURL      \undefined \def \showURL       {\relax}        \fi
\providecommand\bibfield[2]{#2}
\providecommand\bibinfo[2]{#2}
\providecommand\natexlab[1]{#1}
\providecommand\showeprint[2][]{arXiv:#2}

\bibitem[Ablamowicz and Fauser(2007)]%
        {Ablamowicz07}
\bibfield{author}{\bibinfo{person}{Rafal Ablamowicz} {and}
  \bibinfo{person}{Bertfried Fauser}.} \bibinfo{year}{2007}\natexlab{}.
\newblock \bibinfo{booktitle}{\emph{CLIFFORD: a Maple 11 Package for Clifford
  Algebra Computations, version 11}}.
\newblock
\urldef\tempurl%
\url{http://math.tntech.edu/rafal/cliff11/index.html}
\showURL{%
Retrieved February 28, 2008 from \tempurl}


\bibitem[Abril and Plant(2007)]%
        {Abril07}
\bibfield{author}{\bibinfo{person}{Patricia~S. Abril} {and}
  \bibinfo{person}{Robert Plant}.} \bibinfo{year}{2007}\natexlab{}.
\newblock \showarticletitle{The patent holder's dilemma: Buy, sell, or troll?}
\newblock \bibinfo{journal}{\emph{Commun. ACM}} \bibinfo{volume}{50},
  \bibinfo{number}{1} (\bibinfo{date}{Jan.} \bibinfo{year}{2007}),
  \bibinfo{pages}{36--44}.
\newblock
\urldef\tempurl%
\url{https://doi.org/10.1145/1188913.1188915}
\showDOI{\tempurl}


\bibitem[Andler(1979)]%
        {Andler79}
\bibfield{author}{\bibinfo{person}{Sten Andler}.}
  \bibinfo{year}{1979}\natexlab{}.
\newblock \showarticletitle{Predicate Path expressions}. In
  \bibinfo{booktitle}{\emph{Proceedings of the 6th. ACM SIGACT-SIGPLAN
  symposium on Principles of Programming Languages}}
  \emph{(\bibinfo{series}{POPL '79})}. \bibinfo{publisher}{ACM Press},
  \bibinfo{address}{New York, NY}, \bibinfo{pages}{226--236}.
\newblock
\urldef\tempurl%
\url{https://doi.org/10.1145/567752.567774}
\showDOI{\tempurl}


\bibitem[Anisi(2003)]%
        {anisi03}
\bibfield{author}{\bibinfo{person}{David~A. Anisi}.}
  \bibinfo{year}{2003}\natexlab{}.
\newblock \emph{\bibinfo{title}{Optimal Motion Control of a Ground Vehicle}}.
\newblock \bibinfo{thesistype}{Master's\ thesis}. \bibinfo{school}{Royal
  Institute of Technology (KTH), Stockholm, Sweden}.
\newblock


\bibitem[Anzaroot and McCallum(2013)]%
        {UMassCitations}
\bibfield{author}{\bibinfo{person}{Sam Anzaroot} {and} \bibinfo{person}{Andrew
  McCallum}.} \bibinfo{year}{2013}\natexlab{}.
\newblock \bibinfo{booktitle}{\emph{{UMass} Citation Field Extraction
  Dataset}}.
\newblock
\urldef\tempurl%
\url{http://www.iesl.cs.umass.edu/data/data-umasscitationfield}
\showURL{%
Retrieved May 27, 2019 from \tempurl}


\bibitem[Anzaroot et~al\mbox{.}(2014)]%
        {AnzarootPBM14}
\bibfield{author}{\bibinfo{person}{Sam Anzaroot}, \bibinfo{person}{Alexandre
  Passos}, \bibinfo{person}{David Belanger}, {and} \bibinfo{person}{Andrew
  McCallum}.} \bibinfo{year}{2014}\natexlab{}.
\newblock \bibinfo{title}{Learning Soft Linear Constraints with Application to
  Citation Field Extraction}.
\newblock
\newblock
\showeprint[arxiv]{1403.1349}


\bibitem[Bornmann et~al\mbox{.}(2019)]%
        {Bornmann2019}
\bibfield{author}{\bibinfo{person}{Lutz Bornmann}, \bibinfo{person}{K.~Brad
  Wray}, {and} \bibinfo{person}{Robin Haunschild}.}
  \bibinfo{year}{2019}\natexlab{}.
\newblock \bibinfo{title}{Citation concept analysis {(CCA)}---A new form of
  citation analysis revealing the usefulness of concepts for other researchers
  illustrated by two exemplary case studies including classic books by {Thomas
  S.~Kuhn} and {Karl R.~Popper}}.
\newblock
\newblock
\showeprint[arxiv]{1905.12410}~[cs.DL]


\bibitem[Clarkson(1985)]%
        {Clarkson85}
\bibfield{author}{\bibinfo{person}{Kenneth~L. Clarkson}.}
  \bibinfo{year}{1985}\natexlab{}.
\newblock \emph{\bibinfo{title}{Algorithms for Closest-Point Problems
  (Computational Geometry)}}.
\newblock \bibinfo{thesistype}{Ph.\,D. Dissertation}. \bibinfo{school}{Stanford
  University}, \bibinfo{address}{Palo Alto, CA}.
\newblock
\newblock
\shownote{UMI Order Number: AAT 8506171}.


\bibitem[Cohen(1996)]%
        {JCohen96}
\bibfield{editor}{\bibinfo{person}{Jacques Cohen}} (Ed.).
  \bibinfo{year}{1996}\natexlab{}. \showarticletitle{Special issue: Digital
  Libraries}.
\newblock \bibinfo{journal}{\emph{Commun. {ACM}}} \bibinfo{volume}{39},
  \bibinfo{number}{11} (\bibinfo{date}{Nov.} \bibinfo{year}{1996}).

\bibitem[Cohen et~al\mbox{.}(2007)]%
        {Cohen07}
\bibfield{author}{\bibinfo{person}{Sarah Cohen}, \bibinfo{person}{Werner Nutt},
  {and} \bibinfo{person}{Yehoshua Sagic}.} \bibinfo{year}{2007}\natexlab{}.
\newblock \showarticletitle{Deciding equivalances among conjunctive aggregate
  queries}.
\newblock \bibinfo{journal}{\emph{J. ACM}} \bibinfo{volume}{54},
  \bibinfo{number}{2}, Article \bibinfo{articleno}{5} (\bibinfo{date}{April}
  \bibinfo{year}{2007}), \bibinfo{numpages}{50}~pages.
\newblock
\urldef\tempurl%
\url{https://doi.org/10.1145/1219092.1219093}
\showDOI{\tempurl}


\bibitem[Douglass et~al\mbox{.}(1998)]%
        {Douglass98}
\bibfield{author}{\bibinfo{person}{Bruce~P. Douglass}, \bibinfo{person}{David
  Harel}, {and} \bibinfo{person}{Mark~B. Trakhtenbrot}.}
  \bibinfo{year}{1998}\natexlab{}.
\newblock \showarticletitle{Statecarts in use: structured analysis and
  object-orientation}.
\newblock In \bibinfo{booktitle}{\emph{Lectures on Embedded Systems}},
  \bibfield{editor}{\bibinfo{person}{Grzegorz Rozenberg} {and}
  \bibinfo{person}{Frits~W. Vaandrager}} (Eds.). \bibinfo{series}{Lecture Notes
  in Computer Science}, Vol.~\bibinfo{volume}{1494}.
  \bibinfo{publisher}{Springer-Verlag}, \bibinfo{address}{London},
  \bibinfo{pages}{368--394}.
\newblock
\urldef\tempurl%
\url{https://doi.org/10.1007/3-540-65193-4_29}
\showDOI{\tempurl}


\bibitem[Editor(2007)]%
        {Editor00}
\bibfield{editor}{\bibinfo{person}{Ian Editor}} (Ed.).
  \bibinfo{year}{2007}\natexlab{}.
\newblock \bibinfo{booktitle}{\emph{The title of book one}
  (\bibinfo{edition}{1st.} ed.)}. \bibinfo{series}{The name of the series one},
  Vol.~\bibinfo{volume}{9}.
\newblock \bibinfo{publisher}{University of Chicago Press},
  \bibinfo{address}{Chicago}.
\newblock
\urldef\tempurl%
\url{https://doi.org/10.1007/3-540-09237-4}
\showDOI{\tempurl}


\bibitem[Editor(2008)]%
        {Editor00a}
\bibfield{editor}{\bibinfo{person}{Ian Editor}} (Ed.).
  \bibinfo{year}{2008}\natexlab{}.
\newblock \bibinfo{booktitle}{\emph{The title of book two}
  (\bibinfo{edition}{2nd.} ed.)}.
\newblock \bibinfo{publisher}{University of Chicago Press},
  \bibinfo{address}{Chicago}, Chapter 100.
\newblock
\urldef\tempurl%
\url{https://doi.org/10.1007/3-540-09237-4}
\showDOI{\tempurl}


\bibitem[Gundy et~al\mbox{.}(2007)]%
        {VanGundy07}
\bibfield{author}{\bibinfo{person}{Matthew~Van Gundy}, \bibinfo{person}{Davide
  Balzarotti}, {and} \bibinfo{person}{Giovanni Vigna}.}
  \bibinfo{year}{2007}\natexlab{}.
\newblock \showarticletitle{Catch me, if you can: Evading network signatures
  with web-based polymorphic worms}. In \bibinfo{booktitle}{\emph{Proceedings
  of the first USENIX workshop on Offensive Technologies}}
  \emph{(\bibinfo{series}{WOOT '07})}. \bibinfo{publisher}{USENIX Association},
  \bibinfo{address}{Berkley, CA}, Article \bibinfo{articleno}{7},
  \bibinfo{numpages}{9}~pages.
\newblock


\bibitem[Hagerup et~al\mbox{.}(1993)]%
        {Hagerup1993}
\bibfield{author}{\bibinfo{person}{Torben Hagerup}, \bibinfo{person}{Kurt
  Mehlhorn}, {and} \bibinfo{person}{J.~Ian Munro}.}
  \bibinfo{year}{1993}\natexlab{}.
\newblock \showarticletitle{Maintaining Discrete Probability Distributions
  Optimally}. In \bibinfo{booktitle}{\emph{Proceedings of the 20th
  International Colloquium on Automata, Languages and Programming}}
  \emph{(\bibinfo{series}{Lecture Notes in Computer Science},
  Vol.~\bibinfo{volume}{700})}. \bibinfo{publisher}{Springer-Verlag},
  \bibinfo{address}{Berlin}, \bibinfo{pages}{253--264}.
\newblock


\bibitem[Harel(1978)]%
        {Harel78}
\bibfield{author}{\bibinfo{person}{David Harel}.}
  \bibinfo{year}{1978}\natexlab{}.
\newblock \bibinfo{booktitle}{\emph{LOGICS of Programs: AXIOMATICS and
  DESCRIPTIVE POWER}}.
\newblock \bibinfo{type}{MIT Research Lab Technical Report} TR-200.
  \bibinfo{institution}{Massachusetts Institute of Technology},
  \bibinfo{address}{Cambridge, MA}.
\newblock


\bibitem[Harel(1979)]%
        {Harel79}
\bibfield{author}{\bibinfo{person}{David Harel}.}
  \bibinfo{year}{1979}\natexlab{}.
\newblock \bibinfo{booktitle}{\emph{First-Order Dynamic Logic}}.
  \bibinfo{series}{Lecture Notes in Computer Science},
  Vol.~\bibinfo{volume}{68}.
\newblock \bibinfo{publisher}{Springer-Verlag}, \bibinfo{address}{New York,
  NY}.
\newblock
\urldef\tempurl%
\url{https://doi.org/10.1007/3-540-09237-4}
\showDOI{\tempurl}


\bibitem[H{\"o}rmander(1985a)]%
        {MR781537}
\bibfield{author}{\bibinfo{person}{Lars H{\"o}rmander}.}
  \bibinfo{year}{1985}\natexlab{a}.
\newblock \bibinfo{booktitle}{\emph{The analysis of linear partial differential
  operators. {III}}}. \bibinfo{series}{Grundlehren der Mathematischen
  Wissenschaften [Fundamental Principles of Mathematical Sciences]},
  Vol.~\bibinfo{volume}{275}.
\newblock \bibinfo{publisher}{Springer-Verlag}, \bibinfo{address}{Berlin,
  Germany}. viii+525 pages.
\newblock
\showISBNx{3-540-13828-5}
\newblock
\shownote{Pseudodifferential operators}.


\bibitem[H{\"o}rmander(1985b)]%
        {MR781536}
\bibfield{author}{\bibinfo{person}{Lars H{\"o}rmander}.}
  \bibinfo{year}{1985}\natexlab{b}.
\newblock \bibinfo{booktitle}{\emph{The analysis of linear partial differential
  operators. {IV}}}. \bibinfo{series}{Grundlehren der Mathematischen
  Wissenschaften [Fundamental Principles of Mathematical Sciences]},
  Vol.~\bibinfo{volume}{275}.
\newblock \bibinfo{publisher}{Springer-Verlag}, \bibinfo{address}{Berlin,
  Germany}. vii+352 pages.
\newblock
\showISBNx{3-540-13829-3}
\newblock
\shownote{Fourier integral operators}.


\bibitem[IEEE(2004)]%
        {2004:ITE:1009386.1010128}
IEEE \bibinfo{year}{2004}\natexlab{}.
\newblock \showarticletitle{IEEE TCSC Executive Committee}. In
  \bibinfo{booktitle}{\emph{Proceedings of the IEEE International Conference on
  Web Services}} \emph{(\bibinfo{series}{ICWS '04})}. \bibinfo{publisher}{IEEE
  Computer Society}, \bibinfo{address}{Washington, DC, USA},
  \bibinfo{pages}{21--22}.
\newblock
\showISBNx{0-7695-2167-3}
\urldef\tempurl%
\url{https://doi.org/10.1109/ICWS.2004.64}
\showDOI{\tempurl}


\bibitem[Kirschmer and Voight(2010)]%
        {Kirschmer:2010:AEI:1958016.1958018}
\bibfield{author}{\bibinfo{person}{Markus Kirschmer} {and}
  \bibinfo{person}{John Voight}.} \bibinfo{year}{2010}\natexlab{}.
\newblock \showarticletitle{Algorithmic Enumeration of Ideal Classes for
  Quaternion Orders}.
\newblock \bibinfo{journal}{\emph{SIAM J. Comput.}} \bibinfo{volume}{39},
  \bibinfo{number}{5} (\bibinfo{date}{Jan.} \bibinfo{year}{2010}),
  \bibinfo{pages}{1714--1747}.
\newblock
\showISSN{0097-5397}
\urldef\tempurl%
\url{https://doi.org/10.1137/080734467}
\showDOI{\tempurl}


\bibitem[Knuth(1997)]%
        {Knuth97}
\bibfield{author}{\bibinfo{person}{Donald~E. Knuth}.}
  \bibinfo{year}{1997}\natexlab{}.
\newblock \bibinfo{booktitle}{\emph{The Art of Computer Programming, Vol. 1:
  Fundamental Algorithms (3rd. ed.)}}.
\newblock \bibinfo{publisher}{Addison Wesley Longman Publishing Co., Inc.}
\newblock


\bibitem[Kosiur(2001)]%
        {Kosiur01}
\bibfield{author}{\bibinfo{person}{David Kosiur}.}
  \bibinfo{year}{2001}\natexlab{}.
\newblock \bibinfo{booktitle}{\emph{Understanding Policy-Based Networking}
  (\bibinfo{edition}{2nd.} ed.)}.
\newblock \bibinfo{publisher}{Wiley}, \bibinfo{address}{New York, NY}.
\newblock


\bibitem[Lamport(1986)]%
        {Lamport:LaTeX}
\bibfield{author}{\bibinfo{person}{Leslie Lamport}.}
  \bibinfo{year}{1986}\natexlab{}.
\newblock \bibinfo{booktitle}{\emph{\it {\LaTeX}: A Document Preparation
  System}}.
\newblock \bibinfo{publisher}{Addison-Wesley}, \bibinfo{address}{Reading, MA.}
\newblock


\bibitem[Lee(2005)]%
        {Lee05}
\bibfield{author}{\bibinfo{person}{Newton Lee}.}
  \bibinfo{year}{2005}\natexlab{}.
\newblock \showarticletitle{Interview with Bill Kinder: January 13, 2005}.
\newblock \bibinfo{howpublished}{Video}.
\newblock \bibinfo{journal}{\emph{Comput. Entertain.}} \bibinfo{volume}{3},
  \bibinfo{number}{1}, Article \bibinfo{articleno}{4}
  (\bibinfo{date}{Jan.-March} \bibinfo{year}{2005}).
\newblock
\urldef\tempurl%
\url{https://doi.org/10.1145/1057270.1057278}
\showDOI{\tempurl}


\bibitem[Novak(2003)]%
        {Novak03}
\bibfield{author}{\bibinfo{person}{Dave Novak}.}
  \bibinfo{year}{2003}\natexlab{}.
\newblock \showarticletitle{Solder man}. \bibinfo{howpublished}{Video}. In
  \bibinfo{booktitle}{\emph{ACM SIGGRAPH 2003 Video Review on Animation theater
  Program: Part I - Vol. 145 (July 27--27, 2003)}}. \bibinfo{publisher}{ACM
  Press}, \bibinfo{address}{New York, NY}, \bibinfo{pages}{4}.
\newblock
\urldef\tempurl%
\url{https://doi.org/99.9999/woot07-S422}
\showDOI{\tempurl}
\urldef\tempurl%
\url{http://video.google.com/videoplay?docid=6528042696351994555}
\showURL{%
\tempurl}


\bibitem[Obama(2008)]%
        {Obama08}
\bibfield{author}{\bibinfo{person}{Barack Obama}.}
  \bibinfo{year}{2008}\natexlab{}.
\newblock \bibinfo{title}{A more perfect union}.
\newblock \bibinfo{howpublished}{Video}.
\newblock
\urldef\tempurl%
\url{http://video.google.com/videoplay?docid=6528042696351994555}
\showURL{%
Retrieved March 21, 2008 from \tempurl}


\bibitem[Poker-Edge.Com(2006)]%
        {Poker06}
\bibfield{author}{\bibinfo{person}{Poker-Edge.Com}.}
  \bibinfo{year}{2006}\natexlab{}.
\newblock \bibinfo{title}{Stats and Analysis}.
\newblock
\newblock
\urldef\tempurl%
\url{http://www.poker-edge.com/stats.php}
\showURL{%
Retrieved June 7, 2006 from \tempurl}


\bibitem[{R Core Team}(2019)]%
        {R}
\bibfield{author}{\bibinfo{person}{{R Core Team}}.}
  \bibinfo{year}{2019}\natexlab{}.
\newblock \bibinfo{booktitle}{\emph{R: A Language and Environment for
  Statistical Computing}}.
\newblock R Foundation for Statistical Computing, Vienna, Austria.
\newblock
\urldef\tempurl%
\url{https://www.R-project.org/}
\showURL{%
\tempurl}


\bibitem[Rous(2008)]%
        {rous08}
\bibfield{author}{\bibinfo{person}{Bernard Rous}.}
  \bibinfo{year}{2008}\natexlab{}.
\newblock \showarticletitle{The Enabling of Digital Libraries}.
\newblock \bibinfo{journal}{\emph{Digital Libraries}} \bibinfo{volume}{12},
  \bibinfo{number}{3}, Article \bibinfo{articleno}{5} (\bibinfo{date}{July}
  \bibinfo{year}{2008}).
\newblock
\newblock
\shownote{To appear}.


\bibitem[Saeedi et~al\mbox{.}(2010a)]%
        {SaeediMEJ10}
\bibfield{author}{\bibinfo{person}{Mehdi Saeedi},
  \bibinfo{person}{Morteza~Saheb Zamani}, {and} \bibinfo{person}{Mehdi
  Sedighi}.} \bibinfo{year}{2010}\natexlab{a}.
\newblock \showarticletitle{A library-based synthesis methodology for
  reversible logic}.
\newblock \bibinfo{journal}{\emph{Microelectron. J.}} \bibinfo{volume}{41},
  \bibinfo{number}{4} (\bibinfo{date}{April} \bibinfo{year}{2010}),
  \bibinfo{pages}{185--194}.
\newblock


\bibitem[Saeedi et~al\mbox{.}(2010b)]%
        {SaeediJETC10}
\bibfield{author}{\bibinfo{person}{Mehdi Saeedi},
  \bibinfo{person}{Morteza~Saheb Zamani}, \bibinfo{person}{Mehdi Sedighi},
  {and} \bibinfo{person}{Zahra Sasanian}.} \bibinfo{year}{2010}\natexlab{b}.
\newblock \showarticletitle{Synthesis of Reversible Circuit Using Cycle-Based
  Approach}.
\newblock \bibinfo{journal}{\emph{J. Emerg. Technol. Comput. Syst.}}
  \bibinfo{volume}{6}, \bibinfo{number}{4} (\bibinfo{date}{Dec.}
  \bibinfo{year}{2010}).
\newblock


\bibitem[Scientist(2009)]%
        {JoeScientist001}
\bibfield{author}{\bibinfo{person}{Joseph Scientist}.}
  \bibinfo{year}{2009}\natexlab{}.
\newblock \bibinfo{title}{The fountain of youth}.
\newblock
\newblock
\newblock
\shownote{Patent No. 12345, Filed July 1st., 2008, Issued Aug. 9th., 2009}.


\bibitem[Smith(2010)]%
        {Smith10}
\bibfield{author}{\bibinfo{person}{Stan~W. Smith}.}
  \bibinfo{year}{2010}\natexlab{}.
\newblock \showarticletitle{An experiment in bibliographic mark-up: Parsing
  metadata for XML export}. In \bibinfo{booktitle}{\emph{Proceedings of the
  3rd. annual workshop on Librarians and Computers}}
  \emph{(\bibinfo{series}{LAC '10}, Vol.~\bibinfo{volume}{3})},
  \bibfield{editor}{\bibinfo{person}{Reginald~N. Smythe} {and}
  \bibinfo{person}{Alexander Noble}} (Eds.). \bibinfo{publisher}{Paparazzi
  Press}, \bibinfo{address}{Milan Italy}, \bibinfo{pages}{422--431}.
\newblock
\urldef\tempurl%
\url{https://doi.org/99.9999/woot07-S422}
\showDOI{\tempurl}


\bibitem[Spector(1990)]%
        {Spector90}
\bibfield{author}{\bibinfo{person}{Asad~Z. Spector}.}
  \bibinfo{year}{1990}\natexlab{}.
\newblock \showarticletitle{Achieving application requirements}.
\newblock In \bibinfo{booktitle}{\emph{Distributed Systems}
  (\bibinfo{edition}{2nd.} ed.)}, \bibfield{editor}{\bibinfo{person}{Sape
  Mullender}} (Ed.). \bibinfo{publisher}{ACM Press}, \bibinfo{address}{New
  York, NY}, \bibinfo{pages}{19--33}.
\newblock
\urldef\tempurl%
\url{https://doi.org/10.1145/90417.90738}
\showDOI{\tempurl}


\bibitem[Thornburg(2001)]%
        {Thornburg01}
\bibfield{author}{\bibinfo{person}{Harry Thornburg}.}
  \bibinfo{year}{2001}\natexlab{}.
\newblock \bibinfo{booktitle}{\emph{Introduction to Bayesian Statistics}}.
\newblock
\urldef\tempurl%
\url{http://ccrma.stanford.edu/~jos/bayes/bayes.html}
\showURL{%
Retrieved March 2, 2005 from \tempurl}


\bibitem[TUG(2017)]%
        {TUGInstmem}
TUG \bibinfo{year}{2017}\natexlab{}.
\newblock \bibinfo{booktitle}{\emph{Institutional members of the {\TeX} Users
  Group}}.
\newblock
\urldef\tempurl%
\url{http://wwtug.org/instmem.html}
\showURL{%
Retrieved May 27, 2017 from \tempurl}


\bibitem[Veytsman(2017)]%
        {CTANacmart}
\bibfield{author}{\bibinfo{person}{Boris Veytsman}.}
  \bibinfo{year}{2017}\natexlab{}.
\newblock \bibinfo{booktitle}{\emph{acmart---{C}lass for typesetting
  publications of {ACM}}}.
\newblock
\urldef\tempurl%
\url{http://www.ctan.org/pkg/acmart}
\showURL{%
Retrieved May 27, 2017 from \tempurl}


\end{thebibliography}



\begin{thebibliography}{48}


\ifx \showCODEN    \undefined \def \showCODEN     #1{\unskip}     \fi
\ifx \showDOI      \undefined \def \showDOI       #1{#1}\fi
\ifx \showISBNx    \undefined \def \showISBNx     #1{\unskip}     \fi
\ifx \showISBNxiii \undefined \def \showISBNxiii  #1{\unskip}     \fi
\ifx \showISSN     \undefined \def \showISSN      #1{\unskip}     \fi
\ifx \showLCCN     \undefined \def \showLCCN      #1{\unskip}     \fi
\ifx \shownote     \undefined \def \shownote      #1{#1}          \fi
\ifx \showarticletitle \undefined \def \showarticletitle #1{#1}   \fi
\ifx \showURL      \undefined \def \showURL       {\relax}        \fi
\providecommand\bibfield[2]{#2}
\providecommand\bibinfo[2]{#2}
\providecommand\natexlab[1]{#1}
\providecommand\showeprint[2][]{arXiv:#2}

\bibitem[Abbasi-Yadkori et~al\mbox{.}(2011)]%
        {abbasi2011improved}
\bibfield{author}{\bibinfo{person}{Yasin Abbasi-Yadkori}, \bibinfo{person}{D{\'a}vid P{\'a}l}, {and} \bibinfo{person}{Csaba Szepesv{\'a}ri}.} \bibinfo{year}{2011}\natexlab{}.
\newblock \showarticletitle{Improved algorithms for linear stochastic bandits}.
\newblock \bibinfo{journal}{\emph{Advances in neural information processing systems}}  \bibinfo{volume}{24} (\bibinfo{year}{2011}).
\newblock


\bibitem[Agarwal et~al\mbox{.}(2014)]%
        {agarwal2014taming}
\bibfield{author}{\bibinfo{person}{Alekh Agarwal}, \bibinfo{person}{Daniel Hsu}, \bibinfo{person}{Satyen Kale}, \bibinfo{person}{John Langford}, \bibinfo{person}{Lihong Li}, {and} \bibinfo{person}{Robert Schapire}.} \bibinfo{year}{2014}\natexlab{}.
\newblock \showarticletitle{Taming the monster: A fast and simple algorithm for contextual bandits}. In \bibinfo{booktitle}{\emph{International Conference on Machine Learning}}. PMLR, \bibinfo{pages}{1638--1646}.
\newblock


\bibitem[Agrawal and Goyal(2013)]%
        {agrawal2013thompson}
\bibfield{author}{\bibinfo{person}{Shipra Agrawal} {and} \bibinfo{person}{Navin Goyal}.} \bibinfo{year}{2013}\natexlab{}.
\newblock \showarticletitle{Thompson sampling for contextual bandits with linear payoffs}. In \bibinfo{booktitle}{\emph{International conference on machine learning}}. PMLR, \bibinfo{pages}{127--135}.
\newblock


\bibitem[Aharon et~al\mbox{.}(2015)]%
        {aharon2015excuseme}
\bibfield{author}{\bibinfo{person}{Michal Aharon}, \bibinfo{person}{Oren Anava}, \bibinfo{person}{Noa Avigdor-Elgrabli}, \bibinfo{person}{Dana Drachsler-Cohen}, \bibinfo{person}{Shahar Golan}, {and} \bibinfo{person}{Oren Somekh}.} \bibinfo{year}{2015}\natexlab{}.
\newblock \showarticletitle{Excuseme: Asking users to help in item cold-start recommendations}. In \bibinfo{booktitle}{\emph{Proceedings of the 9th ACM Conference on Recommender Systems}}. \bibinfo{pages}{83--90}.
\newblock


\bibitem[Auer et~al\mbox{.}(2002)]%
        {auer2002finite}
\bibfield{author}{\bibinfo{person}{Peter Auer}, \bibinfo{person}{Nicolo Cesa-Bianchi}, {and} \bibinfo{person}{Paul Fischer}.} \bibinfo{year}{2002}\natexlab{}.
\newblock \showarticletitle{Finite-time analysis of the multiarmed bandit problem}.
\newblock \bibinfo{journal}{\emph{Machine learning}} \bibinfo{volume}{47}, \bibinfo{number}{2} (\bibinfo{year}{2002}), \bibinfo{pages}{235--256}.
\newblock


\bibitem[Bajari et~al\mbox{.}(2021)]%
        {bajari2021multiple}
\bibfield{author}{\bibinfo{person}{Patrick Bajari}, \bibinfo{person}{Brian Burdick}, \bibinfo{person}{Guido~W Imbens}, \bibinfo{person}{Lorenzo Masoero}, \bibinfo{person}{James McQueen}, \bibinfo{person}{Thomas Richardson}, {and} \bibinfo{person}{Ido~M Rosen}.} \bibinfo{year}{2021}\natexlab{}.
\newblock \showarticletitle{Multiple randomization designs}.
\newblock \bibinfo{journal}{\emph{arXiv preprint arXiv:2112.13495}} (\bibinfo{year}{2021}).
\newblock


\bibitem[Bendada et~al\mbox{.}(2020)]%
        {bendada2020carousel}
\bibfield{author}{\bibinfo{person}{Walid Bendada}, \bibinfo{person}{Guillaume Salha}, {and} \bibinfo{person}{Th{\'e}o Bontempelli}.} \bibinfo{year}{2020}\natexlab{}.
\newblock \showarticletitle{Carousel personalization in music streaming apps with contextual bandits}. In \bibinfo{booktitle}{\emph{Proceedings of the 14th ACM Conference on Recommender Systems}}. \bibinfo{pages}{420--425}.
\newblock


\bibitem[Chapelle and Li(2011)]%
        {chapelle2011empirical}
\bibfield{author}{\bibinfo{person}{Olivier Chapelle} {and} \bibinfo{person}{Lihong Li}.} \bibinfo{year}{2011}\natexlab{}.
\newblock \showarticletitle{An empirical evaluation of thompson sampling}.
\newblock \bibinfo{journal}{\emph{Advances in neural information processing systems}}  \bibinfo{volume}{24} (\bibinfo{year}{2011}).
\newblock


\bibitem[Chen(2021)]%
        {chen2021exploration}
\bibfield{author}{\bibinfo{person}{Minmin Chen}.} \bibinfo{year}{2021}\natexlab{}.
\newblock \showarticletitle{Exploration in recommender systems}. In \bibinfo{booktitle}{\emph{Fifteenth ACM Conference on Recommender Systems}}. \bibinfo{pages}{551--553}.
\newblock


\bibitem[Chen et~al\mbox{.}(2019)]%
        {chen2019top}
\bibfield{author}{\bibinfo{person}{Minmin Chen}, \bibinfo{person}{Alex Beutel}, \bibinfo{person}{Paul Covington}, \bibinfo{person}{Sagar Jain}, \bibinfo{person}{Francois Belletti}, {and} \bibinfo{person}{Ed~H Chi}.} \bibinfo{year}{2019}\natexlab{}.
\newblock \showarticletitle{Top-k off-policy correction for a REINFORCE recommender system}. In \bibinfo{booktitle}{\emph{Proceedings of the Twelfth ACM International Conference on Web Search and Data Mining}}. \bibinfo{pages}{456--464}.
\newblock


\bibitem[Chen et~al\mbox{.}(2021)]%
        {chen2021values}
\bibfield{author}{\bibinfo{person}{Minmin Chen}, \bibinfo{person}{Yuyan Wang}, \bibinfo{person}{Can Xu}, \bibinfo{person}{Ya Le}, \bibinfo{person}{Mohit Sharma}, \bibinfo{person}{Lee Richardson}, \bibinfo{person}{Su-Lin Wu}, {and} \bibinfo{person}{Ed Chi}.} \bibinfo{year}{2021}\natexlab{}.
\newblock \showarticletitle{Values of User Exploration in Recommender Systems}. In \bibinfo{booktitle}{\emph{Fifteenth ACM Conference on Recommender Systems}}. \bibinfo{pages}{85--95}.
\newblock


\bibitem[Cheung et~al\mbox{.}(2019)]%
        {cheung2019thompson}
\bibfield{author}{\bibinfo{person}{Wang~Chi Cheung}, \bibinfo{person}{Vincent Tan}, {and} \bibinfo{person}{Zixin Zhong}.} \bibinfo{year}{2019}\natexlab{}.
\newblock \showarticletitle{A Thompson sampling algorithm for cascading bandits}. In \bibinfo{booktitle}{\emph{The 22nd International Conference on Artificial Intelligence and Statistics}}. PMLR, \bibinfo{pages}{438--447}.
\newblock


\bibitem[Chu et~al\mbox{.}(2011)]%
        {chu2011contextual}
\bibfield{author}{\bibinfo{person}{Wei Chu}, \bibinfo{person}{Lihong Li}, \bibinfo{person}{Lev Reyzin}, {and} \bibinfo{person}{Robert Schapire}.} \bibinfo{year}{2011}\natexlab{}.
\newblock \showarticletitle{Contextual bandits with linear payoff functions}. In \bibinfo{booktitle}{\emph{Proceedings of the Fourteenth International Conference on Artificial Intelligence and Statistics}}. JMLR Workshop and Conference Proceedings, \bibinfo{pages}{208--214}.
\newblock


\bibitem[Covington et~al\mbox{.}(2016)]%
        {covington2016deep}
\bibfield{author}{\bibinfo{person}{Paul Covington}, \bibinfo{person}{Jay Adams}, {and} \bibinfo{person}{Emre Sargin}.} \bibinfo{year}{2016}\natexlab{}.
\newblock \showarticletitle{Deep neural networks for youtube recommendations}. In \bibinfo{booktitle}{\emph{Proceedings of the 10th ACM conference on recommender systems}}. \bibinfo{pages}{191--198}.
\newblock


\bibitem[Ding et~al\mbox{.}(2021)]%
        {ding2021efficient}
\bibfield{author}{\bibinfo{person}{Qin Ding}, \bibinfo{person}{Cho-Jui Hsieh}, {and} \bibinfo{person}{James Sharpnack}.} \bibinfo{year}{2021}\natexlab{}.
\newblock \showarticletitle{An efficient algorithm for generalized linear bandit: Online stochastic gradient descent and thompson sampling}. In \bibinfo{booktitle}{\emph{International Conference on Artificial Intelligence and Statistics}}. PMLR, \bibinfo{pages}{1585--1593}.
\newblock


\bibitem[Durand et~al\mbox{.}(2018)]%
        {durand2018contextual}
\bibfield{author}{\bibinfo{person}{Audrey Durand}, \bibinfo{person}{Charis Achilleos}, \bibinfo{person}{Demetris Iacovides}, \bibinfo{person}{Katerina Strati}, \bibinfo{person}{Georgios~D Mitsis}, {and} \bibinfo{person}{Joelle Pineau}.} \bibinfo{year}{2018}\natexlab{}.
\newblock \showarticletitle{Contextual bandits for adapting treatment in a mouse model of de novo carcinogenesis}. In \bibinfo{booktitle}{\emph{Machine learning for healthcare conference}}. PMLR, \bibinfo{pages}{67--82}.
\newblock


\bibitem[Filippi et~al\mbox{.}(2010)]%
        {filippi2010parametric}
\bibfield{author}{\bibinfo{person}{Sarah Filippi}, \bibinfo{person}{Olivier Cappe}, \bibinfo{person}{Aur{\'e}lien Garivier}, {and} \bibinfo{person}{Csaba Szepesv{\'a}ri}.} \bibinfo{year}{2010}\natexlab{}.
\newblock \showarticletitle{Parametric bandits: The generalized linear case}.
\newblock \bibinfo{journal}{\emph{Advances in Neural Information Processing Systems}}  \bibinfo{volume}{23} (\bibinfo{year}{2010}).
\newblock


\bibitem[Houthooft et~al\mbox{.}(2016)]%
        {houthooft2016vime}
\bibfield{author}{\bibinfo{person}{Rein Houthooft}, \bibinfo{person}{Xi Chen}, \bibinfo{person}{Yan Duan}, \bibinfo{person}{John Schulman}, \bibinfo{person}{Filip De~Turck}, {and} \bibinfo{person}{Pieter Abbeel}.} \bibinfo{year}{2016}\natexlab{}.
\newblock \showarticletitle{Vime: Variational information maximizing exploration}.
\newblock \bibinfo{journal}{\emph{Advances in neural information processing systems}}  \bibinfo{volume}{29} (\bibinfo{year}{2016}).
\newblock


\bibitem[Imbens and Rubin(2015)]%
        {imbens2015causal}
\bibfield{author}{\bibinfo{person}{Guido~W Imbens} {and} \bibinfo{person}{Donald~B Rubin}.} \bibinfo{year}{2015}\natexlab{}.
\newblock \bibinfo{booktitle}{\emph{Causal inference in statistics, social, and biomedical sciences}}.
\newblock \bibinfo{publisher}{Cambridge University Press}.
\newblock


\bibitem[Jadidinejad et~al\mbox{.}(2020)]%
        {jadidinejad2020using}
\bibfield{author}{\bibinfo{person}{Amir~H Jadidinejad}, \bibinfo{person}{Craig Macdonald}, {and} \bibinfo{person}{Iadh Ounis}.} \bibinfo{year}{2020}\natexlab{}.
\newblock \showarticletitle{Using Exploration to Alleviate Closed Loop Effects in Recommender Systems}. In \bibinfo{booktitle}{\emph{Proceedings of the 43rd International ACM SIGIR Conference on Research and Development in Information Retrieval}}. \bibinfo{pages}{2025--2028}.
\newblock


\bibitem[Jeunen and Goethals(2021)]%
        {jeunen2021top}
\bibfield{author}{\bibinfo{person}{Olivier Jeunen} {and} \bibinfo{person}{Bart Goethals}.} \bibinfo{year}{2021}\natexlab{}.
\newblock \showarticletitle{Top-k contextual bandits with equity of exposure}. In \bibinfo{booktitle}{\emph{Proceedings of the 15th ACM Conference on Recommender Systems}}. \bibinfo{pages}{310--320}.
\newblock


\bibitem[Jiang et~al\mbox{.}(2019)]%
        {jiang2019degenerate}
\bibfield{author}{\bibinfo{person}{Ray Jiang}, \bibinfo{person}{Silvia Chiappa}, \bibinfo{person}{Tor Lattimore}, \bibinfo{person}{Andr{\'a}s Gy{\"o}rgy}, {and} \bibinfo{person}{Pushmeet Kohli}.} \bibinfo{year}{2019}\natexlab{}.
\newblock \showarticletitle{Degenerate feedback loops in recommender systems}. In \bibinfo{booktitle}{\emph{Proceedings of the 2019 AAAI/ACM Conference on AI, Ethics, and Society}}. \bibinfo{pages}{383--390}.
\newblock


\bibitem[Joachims et~al\mbox{.}(2018)]%
        {joachims2018deep}
\bibfield{author}{\bibinfo{person}{Thorsten Joachims}, \bibinfo{person}{Adith Swaminathan}, {and} \bibinfo{person}{Maarten De~Rijke}.} \bibinfo{year}{2018}\natexlab{}.
\newblock \showarticletitle{Deep learning with logged bandit feedback}. In \bibinfo{booktitle}{\emph{International Conference on Learning Representations}}.
\newblock


\bibitem[Kohavi et~al\mbox{.}(2020)]%
        {kohavi2020trustworthy}
\bibfield{author}{\bibinfo{person}{Ron Kohavi}, \bibinfo{person}{Diane Tang}, {and} \bibinfo{person}{Ya Xu}.} \bibinfo{year}{2020}\natexlab{}.
\newblock \bibinfo{booktitle}{\emph{Trustworthy online controlled experiments: A practical guide to a/b testing}}.
\newblock \bibinfo{publisher}{Cambridge University Press}.
\newblock


\bibitem[Koren et~al\mbox{.}(2009)]%
        {koren2009matrix}
\bibfield{author}{\bibinfo{person}{Yehuda Koren}, \bibinfo{person}{Robert Bell}, {and} \bibinfo{person}{Chris Volinsky}.} \bibinfo{year}{2009}\natexlab{}.
\newblock \showarticletitle{Matrix factorization techniques for recommender systems}.
\newblock \bibinfo{journal}{\emph{Computer}} \bibinfo{volume}{42}, \bibinfo{number}{8} (\bibinfo{year}{2009}), \bibinfo{pages}{30--37}.
\newblock


\bibitem[Krishnamoorthy and Menon(2013)]%
        {krishnamoorthy2013matrix}
\bibfield{author}{\bibinfo{person}{Aravindh Krishnamoorthy} {and} \bibinfo{person}{Deepak Menon}.} \bibinfo{year}{2013}\natexlab{}.
\newblock \showarticletitle{Matrix inversion using Cholesky decomposition}. In \bibinfo{booktitle}{\emph{2013 signal processing: Algorithms, architectures, arrangements, and applications (SPA)}}. IEEE, \bibinfo{pages}{70--72}.
\newblock


\bibitem[Kveton et~al\mbox{.}(2022)]%
        {kveton2022value}
\bibfield{author}{\bibinfo{person}{Branislav Kveton}, \bibinfo{person}{Ofer Meshi}, \bibinfo{person}{Masrour Zoghi}, {and} \bibinfo{person}{Zhen Qin}.} \bibinfo{year}{2022}\natexlab{}.
\newblock \showarticletitle{On the Value of Prior in Online Learning to Rank}. In \bibinfo{booktitle}{\emph{International Conference on Artificial Intelligence and Statistics}}. PMLR, \bibinfo{pages}{6880--6892}.
\newblock


\bibitem[Kveton et~al\mbox{.}(2020)]%
        {kveton2020randomized}
\bibfield{author}{\bibinfo{person}{Branislav Kveton}, \bibinfo{person}{Manzil Zaheer}, \bibinfo{person}{Csaba Szepesvari}, \bibinfo{person}{Lihong Li}, \bibinfo{person}{Mohammad Ghavamzadeh}, {and} \bibinfo{person}{Craig Boutilier}.} \bibinfo{year}{2020}\natexlab{}.
\newblock \showarticletitle{Randomized exploration in generalized linear bandits}. In \bibinfo{booktitle}{\emph{International Conference on Artificial Intelligence and Statistics}}. PMLR, \bibinfo{pages}{2066--2076}.
\newblock


\bibitem[Lai et~al\mbox{.}(1985)]%
        {lai1985asymptotically}
\bibfield{author}{\bibinfo{person}{Tze~Leung Lai}, \bibinfo{person}{Herbert Robbins}, {et~al\mbox{.}}} \bibinfo{year}{1985}\natexlab{}.
\newblock \showarticletitle{Asymptotically efficient adaptive allocation rules}.
\newblock \bibinfo{journal}{\emph{Advances in applied mathematics}} \bibinfo{volume}{6}, \bibinfo{number}{1} (\bibinfo{year}{1985}), \bibinfo{pages}{4--22}.
\newblock


\bibitem[Langford and Zhang(2007)]%
        {langford2007epoch}
\bibfield{author}{\bibinfo{person}{John Langford} {and} \bibinfo{person}{Tong Zhang}.} \bibinfo{year}{2007}\natexlab{}.
\newblock \showarticletitle{The epoch-greedy algorithm for multi-armed bandits with side information}.
\newblock \bibinfo{journal}{\emph{Advances in neural information processing systems}}  \bibinfo{volume}{20} (\bibinfo{year}{2007}).
\newblock


\bibitem[Li et~al\mbox{.}(2010)]%
        {li2010contextual}
\bibfield{author}{\bibinfo{person}{Lihong Li}, \bibinfo{person}{Wei Chu}, \bibinfo{person}{John Langford}, {and} \bibinfo{person}{Robert~E Schapire}.} \bibinfo{year}{2010}\natexlab{}.
\newblock \showarticletitle{A contextual-bandit approach to personalized news article recommendation}. In \bibinfo{booktitle}{\emph{Proceedings of the 19th international conference on World wide web}}. \bibinfo{pages}{661--670}.
\newblock


\bibitem[Li et~al\mbox{.}(2017)]%
        {li2017provably}
\bibfield{author}{\bibinfo{person}{Lihong Li}, \bibinfo{person}{Yu Lu}, {and} \bibinfo{person}{Dengyong Zhou}.} \bibinfo{year}{2017}\natexlab{}.
\newblock \showarticletitle{Provably optimal algorithms for generalized linear contextual bandits}. In \bibinfo{booktitle}{\emph{International Conference on Machine Learning}}. PMLR, \bibinfo{pages}{2071--2080}.
\newblock


\bibitem[Liu et~al\mbox{.}(2018)]%
        {liu2018customized}
\bibfield{author}{\bibinfo{person}{Bing Liu}, \bibinfo{person}{Tong Yu}, \bibinfo{person}{Ian Lane}, {and} \bibinfo{person}{Ole~J Mengshoel}.} \bibinfo{year}{2018}\natexlab{}.
\newblock \showarticletitle{Customized nonlinear bandits for online response selection in neural conversation models}. In \bibinfo{booktitle}{\emph{Thirty-Second AAAI Conference on Artificial Intelligence}}.
\newblock


\bibitem[McInerney et~al\mbox{.}(2018)]%
        {mcinerney2018explore}
\bibfield{author}{\bibinfo{person}{James McInerney}, \bibinfo{person}{Benjamin Lacker}, \bibinfo{person}{Samantha Hansen}, \bibinfo{person}{Karl Higley}, \bibinfo{person}{Hugues Bouchard}, \bibinfo{person}{Alois Gruson}, {and} \bibinfo{person}{Rishabh Mehrotra}.} \bibinfo{year}{2018}\natexlab{}.
\newblock \showarticletitle{Explore, exploit, and explain: personalizing explainable recommendations with bandits}. In \bibinfo{booktitle}{\emph{Proceedings of the 12th ACM conference on recommender systems}}. \bibinfo{pages}{31--39}.
\newblock


\bibitem[Mintz et~al\mbox{.}(2020)]%
        {mintz2020nonstationary}
\bibfield{author}{\bibinfo{person}{Yonatan Mintz}, \bibinfo{person}{Anil Aswani}, \bibinfo{person}{Philip Kaminsky}, \bibinfo{person}{Elena Flowers}, {and} \bibinfo{person}{Yoshimi Fukuoka}.} \bibinfo{year}{2020}\natexlab{}.
\newblock \showarticletitle{Nonstationary bandits with habituation and recovery dynamics}.
\newblock \bibinfo{journal}{\emph{Operations Research}} \bibinfo{volume}{68}, \bibinfo{number}{5} (\bibinfo{year}{2020}), \bibinfo{pages}{1493--1516}.
\newblock


\bibitem[Misra et~al\mbox{.}(2019)]%
        {misra2019dynamic}
\bibfield{author}{\bibinfo{person}{Kanishka Misra}, \bibinfo{person}{Eric~M Schwartz}, {and} \bibinfo{person}{Jacob Abernethy}.} \bibinfo{year}{2019}\natexlab{}.
\newblock \showarticletitle{Dynamic online pricing with incomplete information using multiarmed bandit experiments}.
\newblock \bibinfo{journal}{\emph{Marketing Science}} \bibinfo{volume}{38}, \bibinfo{number}{2} (\bibinfo{year}{2019}), \bibinfo{pages}{226--252}.
\newblock


\bibitem[Penrose(1955)]%
        {penrose1955generalized}
\bibfield{author}{\bibinfo{person}{Roger Penrose}.} \bibinfo{year}{1955}\natexlab{}.
\newblock \showarticletitle{A generalized inverse for matrices}. In \bibinfo{booktitle}{\emph{Mathematical proceedings of the Cambridge philosophical society}}, Vol.~\bibinfo{volume}{51}. Cambridge University Press, \bibinfo{pages}{406--413}.
\newblock


\bibitem[Press et~al\mbox{.}(2007)]%
        {press2007numerical}
\bibfield{author}{\bibinfo{person}{William~H Press}, \bibinfo{person}{Saul~A Teukolsky}, \bibinfo{person}{William~T Vetterling}, {and} \bibinfo{person}{Brian~P Flannery}.} \bibinfo{year}{2007}\natexlab{}.
\newblock \bibinfo{booktitle}{\emph{Numerical recipes 3rd edition: The art of scientific computing}}.
\newblock \bibinfo{publisher}{Cambridge university press}.
\newblock


\bibitem[Riquelme et~al\mbox{.}(2018)]%
        {riquelme2018deep}
\bibfield{author}{\bibinfo{person}{Carlos Riquelme}, \bibinfo{person}{George Tucker}, {and} \bibinfo{person}{Jasper Snoek}.} \bibinfo{year}{2018}\natexlab{}.
\newblock \showarticletitle{Deep bayesian bandits showdown: An empirical comparison of bayesian deep networks for thompson sampling}.
\newblock \bibinfo{journal}{\emph{arXiv preprint arXiv:1802.09127}} (\bibinfo{year}{2018}).
\newblock


\bibitem[Sarwar et~al\mbox{.}(2001)]%
        {sarwar2001item}
\bibfield{author}{\bibinfo{person}{Badrul Sarwar}, \bibinfo{person}{George Karypis}, \bibinfo{person}{Joseph Konstan}, {and} \bibinfo{person}{John Riedl}.} \bibinfo{year}{2001}\natexlab{}.
\newblock \showarticletitle{Item-based collaborative filtering recommendation algorithms}. In \bibinfo{booktitle}{\emph{Proceedings of the 10th international conference on World Wide Web}}. \bibinfo{pages}{285--295}.
\newblock


\bibitem[Schnabel et~al\mbox{.}(2018)]%
        {schnabel2018short}
\bibfield{author}{\bibinfo{person}{Tobias Schnabel}, \bibinfo{person}{Paul~N Bennett}, \bibinfo{person}{Susan~T Dumais}, {and} \bibinfo{person}{Thorsten Joachims}.} \bibinfo{year}{2018}\natexlab{}.
\newblock \showarticletitle{Short-term satisfaction and long-term coverage: Understanding how users tolerate algorithmic exploration}. In \bibinfo{booktitle}{\emph{Proceedings of the Eleventh ACM International Conference on Web Search and Data Mining}}. \bibinfo{pages}{513--521}.
\newblock


\bibitem[Silver et~al\mbox{.}(2016)]%
        {silver2016mastering}
\bibfield{author}{\bibinfo{person}{David Silver}, \bibinfo{person}{Aja Huang}, \bibinfo{person}{Chris~J Maddison}, \bibinfo{person}{Arthur Guez}, \bibinfo{person}{Laurent Sifre}, \bibinfo{person}{George Van Den~Driessche}, \bibinfo{person}{Julian Schrittwieser}, \bibinfo{person}{Ioannis Antonoglou}, \bibinfo{person}{Veda Panneershelvam}, \bibinfo{person}{Marc Lanctot}, {et~al\mbox{.}}} \bibinfo{year}{2016}\natexlab{}.
\newblock \showarticletitle{Mastering the game of Go with deep neural networks and tree search}.
\newblock \bibinfo{journal}{\emph{nature}} \bibinfo{volume}{529}, \bibinfo{number}{7587} (\bibinfo{year}{2016}), \bibinfo{pages}{484--489}.
\newblock


\bibitem[Song et~al\mbox{.}(2022)]%
        {song2022show}
\bibfield{author}{\bibinfo{person}{Yu Song}, \bibinfo{person}{Shuai Sun}, \bibinfo{person}{Jianxun Lian}, \bibinfo{person}{Hong Huang}, \bibinfo{person}{Yu Li}, \bibinfo{person}{Hai Jin}, {and} \bibinfo{person}{Xing Xie}.} \bibinfo{year}{2022}\natexlab{}.
\newblock \showarticletitle{Show Me the Whole World: Towards Entire Item Space Exploration for Interactive Personalized Recommendations}. In \bibinfo{booktitle}{\emph{Proceedings of the Fifteenth ACM International Conference on Web Search and Data Mining}}. \bibinfo{pages}{947--956}.
\newblock


\bibitem[Thompson(1933)]%
        {thompson1933likelihood}
\bibfield{author}{\bibinfo{person}{William~R Thompson}.} \bibinfo{year}{1933}\natexlab{}.
\newblock \showarticletitle{On the likelihood that one unknown probability exceeds another in view of the evidence of two samples}.
\newblock \bibinfo{journal}{\emph{Biometrika}} \bibinfo{volume}{25}, \bibinfo{number}{3-4} (\bibinfo{year}{1933}), \bibinfo{pages}{285--294}.
\newblock


\bibitem[Zhang et~al\mbox{.}(2019)]%
        {zhang2019deep}
\bibfield{author}{\bibinfo{person}{Shuai Zhang}, \bibinfo{person}{Lina Yao}, \bibinfo{person}{Aixin Sun}, {and} \bibinfo{person}{Yi Tay}.} \bibinfo{year}{2019}\natexlab{}.
\newblock \showarticletitle{Deep learning based recommender system: A survey and new perspectives}.
\newblock \bibinfo{journal}{\emph{ACM Computing Surveys (CSUR)}} \bibinfo{volume}{52}, \bibinfo{number}{1} (\bibinfo{year}{2019}), \bibinfo{pages}{1--38}.
\newblock


\bibitem[Zhang et~al\mbox{.}(2020)]%
        {zhang2020neural}
\bibfield{author}{\bibinfo{person}{Weitong Zhang}, \bibinfo{person}{Dongruo Zhou}, \bibinfo{person}{Lihong Li}, {and} \bibinfo{person}{Quanquan Gu}.} \bibinfo{year}{2020}\natexlab{}.
\newblock \showarticletitle{Neural thompson sampling}.
\newblock \bibinfo{journal}{\emph{arXiv preprint arXiv:2010.00827}} (\bibinfo{year}{2020}).
\newblock


\bibitem[Zhou et~al\mbox{.}(2020)]%
        {zhou2020neural}
\bibfield{author}{\bibinfo{person}{Dongruo Zhou}, \bibinfo{person}{Lihong Li}, {and} \bibinfo{person}{Quanquan Gu}.} \bibinfo{year}{2020}\natexlab{}.
\newblock \showarticletitle{Neural contextual bandits with ucb-based exploration}. In \bibinfo{booktitle}{\emph{International Conference on Machine Learning}}. PMLR, \bibinfo{pages}{11492--11502}.
\newblock


\bibitem[Zong et~al\mbox{.}(2016)]%
        {zong2016cascading}
\bibfield{author}{\bibinfo{person}{Shi Zong}, \bibinfo{person}{Hao Ni}, \bibinfo{person}{Kenny Sung}, \bibinfo{person}{Nan~Rosemary Ke}, \bibinfo{person}{Zheng Wen}, {and} \bibinfo{person}{Branislav Kveton}.} \bibinfo{year}{2016}\natexlab{}.
\newblock \showarticletitle{Cascading bandits for large-scale recommendation problems}.
\newblock \bibinfo{journal}{\emph{arXiv preprint arXiv:1603.05359}} (\bibinfo{year}{2016}).
\newblock


\end{thebibliography}

\newpage
\appendix
\section{Long-term Value of Exploration through Model Learning}
\label{appsec: model}
In this section, we complement our study by examining the benefit of the exploration through another important intermediary: model learning. Besides the benefit from the corpus change, exploration in the uncertain region provides us interaction data with better coverage and less selection bias, facilitating downstream model training and reducing the model uncertainty. We begin with a new experiment design to measure the exploration benefit on model learning, followed by the long-term study on how exploration reduces model uncertainty hence leads to better model quality. 

\subsection{Data-Diverted Experiment}
\label{subsec:data_diverted}
The traditional user-diverted A/B testing fails to measure the gain in the aspect of data and model since: 1) the models in both the control (production system) and the treatment (exploration system) are trained on the same data pool; 2) the exploration traffic in the treatment arm (typically $1\%$ to $5\%$) is often too small to change the training data distribution and consequently model training. 
\begin{figure}[!ht]
    \centering
    \includegraphics[width=0.9\linewidth]{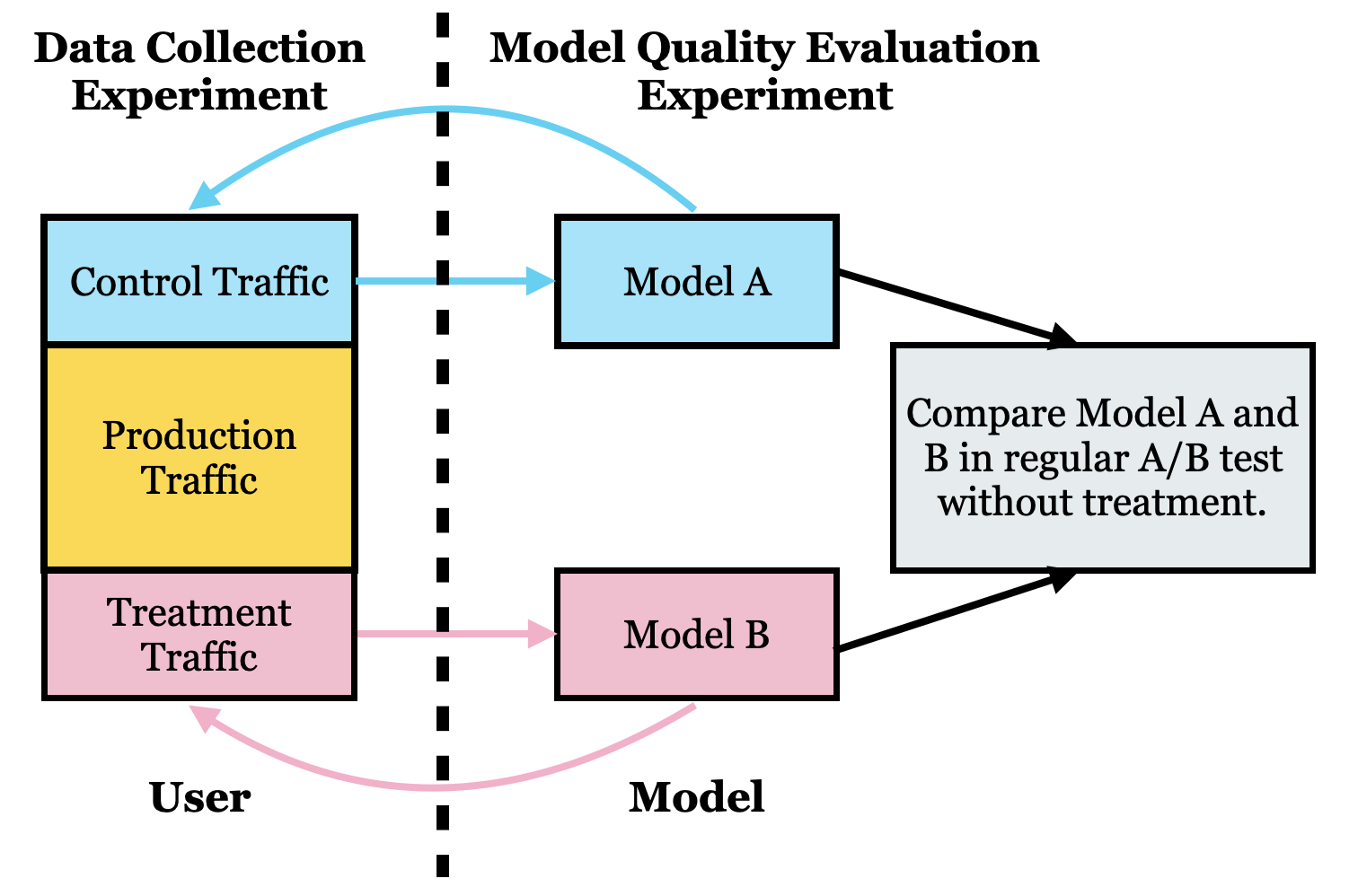}
    \caption{Data-Diverted A/B Testing for measuring exploration benefit on model learning.}
    \label{fig:data_diverted_exp}
\end{figure}

We thus introduce \emph{data-diverted A/B testing}, which separates the logged data from the control and treatment, and compares the quality of the models trained on these disparate data sources. Specifically, the experimental procedure is 
illustrated in Figure~\ref{fig:data_diverted_exp}. We highlight the following three major components: 1) two sets of experiments are running concurrently, one for data collection (Figure~\ref{fig:data_diverted_exp}, left), and one for model quality evaluation (Figure~\ref{fig:data_diverted_exp}, right); 2) the data is logged separately for the control and treatment arm in data collection, and during training the control (treatment) model is trained solely based on the logged data from the control (treatment) group; 3) during serving, we use the same system (in our case, a purely exploitation-based production system) to examine the quality of the models trained on different data sources, eliminating the confounding factor of the algorithmic change (i.e., adding exploration or not).

\subsection{Experiment Setup} 
We then conducted a $5\%$ data-diverted A/B testing where 1) the control arm runs the exploitation-based system as depicted in Figure 1; 2) the treatment arm applies exploration treatment to the ranking system. 
Specifically, it replaces the original \content~ relevance score estimated from the ranker with a sample from the posterior distribution of the relevance score. The uncertainty used in Thompson Sampling is estimated from an ensemble of 5 versions of the same ranking model as in the control arm with different random seeds~\footnote{To reduce training and serving cost, instead of creating 5 copies of the entire model, we shared the bottom representation layers and only created 5 copies of the heads.}. 
\subsection{Long-Term Value of Exploration: Model Uncertainty and Quality}
We study the long-term value of exploration on the learned model from two aspects: (1). reduced model uncertainty, which suggests that the exploration system effectively delves into more uncertain regions and acquires valuable learning signals to reduce uncertainty; 
(2). better model quality: learning signals collected on previously unknown user-\content~ pairs enable the model to more accurately predict \content~ relevance and discover new \content~ that fall into users' interest. 
 
\begin{figure}[!ht]
    \centering
    \includegraphics[width=\linewidth]{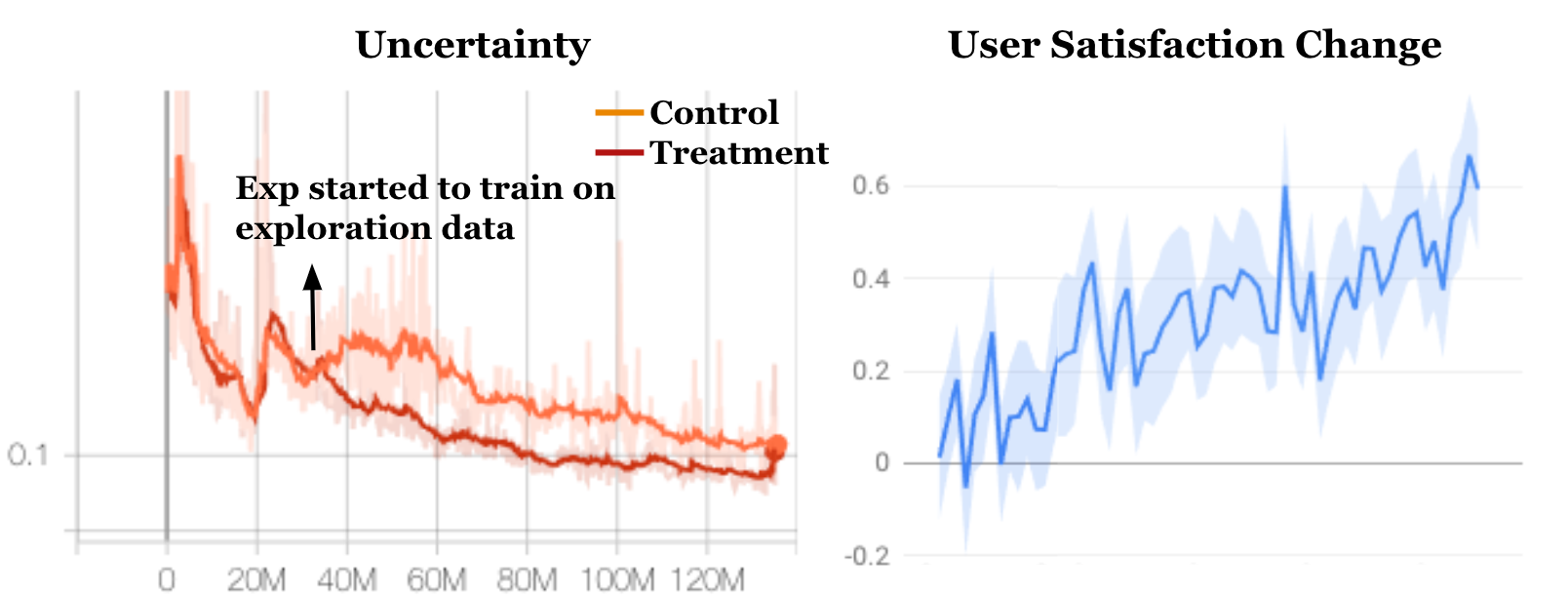}
    \caption{Left: The standard deviation of the relevance score averaged over all the evaluation data, measured from the ensemble of 5 models. Right: The percentage increase between the treatment and control arm, in terms of long-term user satisfaction, over a 3-month period.}
    \label{fig:model_uncertainty}
\end{figure}

\paragraph{Model Uncertainty.} We quantify the uncertainty of the model as the average standard deviation of the relevance prediction $r$ over all (user $u$, item $a$) pairs, from the distribution induced by the evaluation recommendation system $\pi$, i.e., $\EE_{(u,a)\sim \pi}[\sqrt{Var(r(u,a))}]$. Here the evaluation system $\pi$ is the production system used in the data-diverted A/B testing model evaluation stage (figure~\ref{fig:data_diverted_exp} right), and $Var(r(u,a))$ is the uncertainty estimated from the ensemble~\footnote{Note that for both the control and experiment arm we created the ensembles to estimate the variance for model quality evaluation. Only the treatment arm uses the estimated uncertainty in data collection.}. Figure~\ref{fig:model_uncertainty} (left) shows how the empirical model uncertainty changes as the training step increases. First, uncertainty of both the control and experiment model reduces as the training step increases, by observing more data. Second, as the model in the treatment arm starts training using the logged data from the exploration system (around 30M steps), it attains lower model uncertainty compared with the model trained using data from a purely exploitation-based system. In other words, the exploration system helps reduce model uncertainty through deliberate exploration in uncertain regions.  
\paragraph{Model Quality.} We further study how the reduced model uncertainty translates into higher model quality and better long-term user satisfaction. 
In Figure~\ref{fig:model_uncertainty} (right), we plot the percentage increase in user satisfaction between treatment (i.e., exploration-based model) and control. One can easily see that the model trained using the logged data from the exploration-based system leads to significant long-term user satisfaction gain. More importantly, the model quality gain keeps increasing over time (3-month period), demonstrating that the benefit of exploration persists over a long period, and efficient measurement of the exploration gain requires a long-term online experiment.

\section{Corpus Ablation Algorithm}
\label{app:corpus_ablation}
\begin{algorithm}
\SetAlgoLined
\textbf{System:} We use the production multi-stage recommender system as depicted in Figure~\ref{fig:pipeline}. Assume we have $N$ nominators in the first stage, and each nominates $n_i$ candidates initially. \\
\textbf{Ablating the corpus for $x\%$:} upon a request from user $u$,
\begin{itemize}
    \item Each nominator nominates $n^{new}_i = \frac{n_i}{1-x\%}$ candidates $\Acal_i$.
    \item Set the user-specific random seed: $s_{u}$. 
    \item Randomly filter $x\%$ of \contents~ from the output of each nominator $\Acal_i$ by randomly hashing the content id using $s_u$
     and denote the new output as $\Acal^{new}_i$.
    \item Pass $\bigcup_{i=1}^N \Acal^{new}_i$ to the ranking and packing system, which ranks the candidates, and packs a full page and presents to the user.
\end{itemize}
\caption{Corpus Ablation Procedure}
\label{alg:corpus_ablation_procedure}
\end{algorithm}


\end{document}